\documentclass[apj]{emulateapj}

\usepackage{ulem}
\usepackage{psfig}

\begin{document}

\title{Calcium-rich gap transients in the remote outskirts of galaxies}
\author{Mansi M. Kasliwal\altaffilmark{1,2}, S. R. Kulkarni\altaffilmark{1}, 
Avishay Gal-Yam\altaffilmark{3}, Peter E. Nugent\altaffilmark{4,5}, Mark Sullivan\altaffilmark{6}, Lars Bildsten\altaffilmark{7,8}, Ofer Yaron\altaffilmark{3}, Hagai B. Perets\altaffilmark{9}, Iair Arcavi\altaffilmark{3}, Sagi Ben-Ami\altaffilmark{3}, Varun B. Bhalerao\altaffilmark{1}, Joshua S. Bloom\altaffilmark{5}, S. Bradley Cenko\altaffilmark{5}, Alexei V. Filippenko\altaffilmark{5}, Dale A. Frail\altaffilmark{10}, Mohan Ganeshalingam\altaffilmark{5}, Assaf Horesh\altaffilmark{1}, D. Andrew Howell\altaffilmark{8,11}, Nicholas M. Law\altaffilmark{12}, Douglas C. Leonard\altaffilmark{13}, Weidong Li\altaffilmark{5}, Eran O. Ofek\altaffilmark{3}, David Polishook\altaffilmark{3}, Dovi Poznanski\altaffilmark{14}, Robert M. Quimby\altaffilmark{1,15}, Jeffrey M. Silverman\altaffilmark{5}, Assaf Sternberg\altaffilmark{3}, \& Dong Xu\altaffilmark{3}}

\altaffiltext{1}{Cahill Center for Astrophysics, California Institute of Technology, Pasadena, CA, 91125, USA}
\altaffiltext{2}{Observatories of the Carnegie Institution for Science, 813 Santa Barbara St, Pasadena, CA, 91101, USA}
\altaffiltext{3}{Benoziyo Center for Astrophysics, Faculty of Physics, The Weizmann Institute of Science, Rehovot 76100, Israel}
\altaffiltext{4}{Computational Cosmology Center, Lawrence Berkeley National Laboratory, 1 Cyclotron Road, Berkeley, CA 94720, USA}
\altaffiltext{5}{Department of Astronomy, University of California, Berkeley, CA 94720-3411, USA}
\altaffiltext{6}{Department of Physics, Oxford University, Oxford, OX1 3RH, UK}
\altaffiltext{7}{Department of Physics, University of California, Santa Barbara, CA 93106, USA}
\altaffiltext{8}{Kavli Institute for Theoretical Physics, University of California, Santa Barbara, CA 93106, USA}
\altaffiltext{9}{Harvard-Smithsonian Center for Astrophysics, 60 Garden St., Cambridge, MA 02338}
\altaffiltext{10}{National Radio Astronomy Observatory, Array Operations Center, Socorro, NM 87801, USA}
\altaffiltext{11}{Las Cumbres Observatory Global Telescope Network, Inc., Santa Barbara, CA, 93117, USA}
\altaffiltext{12}{Dunlap Institute for Astronomy and Astrophysics, University of Toronto, 50 St. George Street, Toronto M5S 3H4, Ontario, Canada}
\altaffiltext{13}{Department of Astronomy, San Diego State University, San Diego, CA 92182, USA}
\altaffiltext{14}{School of Physics and Astronomy, Tel-Aviv University, Tel Aviv 69978, Israel}
\altaffiltext{15}{IPMU, University of Tokyo, Kashiwanoha 5-1-5, Kashiwa-shi, Chiba, Japan}

\begin{abstract}
From the first two seasons of the Palomar Transient Factory,
we identify three peculiar transients (PTF\,09dav, PTF\,10iuv, PTF\,11bij) 
with five distinguishing characteristics:
peak luminosity in the gap between novae and supernovae ($M_R\,\approx\,-$15.5 to $-$16.5\,mag), 
rapid photometric evolution ($t_{\rm rise}\,\approx\,$12--15\,days), large photospheric velocities 
($\approx$\,6000 to 11000 km s$^{-1}$), early spectroscopic evolution into nebular phase 
($\approx$\,1 to 3\,months) and peculiar nebular spectra dominated by Calcium. 
We also culled the extensive decade-long Lick Observatory Supernova Search database 
and identified an additional member of this group, SN\,2007ke.  Our choice of photometric and
spectroscopic properties was motivated by SN\,2005E \citep{pgm+10}. To our surprise, as in the
case of SN\,2005E, all four members of this group are also clearly offset from the bulk of their host galaxy. 
%Additionally, these transients  
%are found in remote locations, far away (30---40\,kpc) from their putative hosts and 
%have no underlying host brighter than $-$11\,mag. These events are most similar to SN\,2005E \citep{pgm+10}.
Given the well-sampled early and late-time light curves, we derive ejecta masses in the range of
0.4--0.7\,M$_{\odot}$.
% and find it unlikely that the light curve is powered by radioactive $^{56}$Ni. 
Spectroscopically, we find that there may be a diversity in the photospheric phase, but the commonality
is in the unusual nebular spectra. Our extensive follow-up observations rule out standard thermonuclear and standard core-collapse 
explosions for this class of ``Calcium-rich gap'' transients.
If the progenitor is a white dwarf, we are likely seeing a detonation of the white dwarf core
and perhaps even shock-front interaction with a previously ejected nova shell.
In the less likely scenario of a massive star progenitor, 
a nonstandard channel specific to a low-metallicity environment needs to be invoked 
(e.g., ejecta fallback leading to black hole formation). 
Detection (or the lack thereof) of a faint
underlying host (dwarf galaxy, cluster) will provide a
crucial and decisive diagnostic to choose between these alternatives. 
\end{abstract}


\section{Introduction}
\label{sec:introduction}

In the past decade, supernova (SN) surveys targeting nearby, luminous galaxies
have been immensely successful since the total local starlight searched is significantly
larger than in untargeted pointings of equal area. 
%Recently, untargeted wide-angle
%transient surveys have also been successful with the added advantage of finding
%supernovae independent of a host galaxy bias.
The Palomar Transient Factory 
(PTF; \citealt{lkd+09,rkl+09}){\footnote{http://www.astro.caltech.edu/ptf/}} 
is a  wide-angle
transient survey that has an ongoing Dynamic Cadence experiment that combines the 
advantages of both a targeted and an untargeted survey. This experiment searches 
wide-angle pointings of local ($d < 200$\,Mpc) galaxy light 
concentrations at a 1-day cadence to a median depth of 21\,mag. The depth, cadence, and 
choice of pointings allow us to find transients that are fainter, faster, and rarer than supernovae. 
Furthermore, these survey design characteristics facilitate the discovery of 
intracluster transients as well as transients in the far-flung outskirts of their host galaxies. 

While monitoring supernovae from the targeted Lick Observatory Supernova Search \citep[LOSS;][]{lft+00,flt+01} with the Katzman Automatic Imaging Telescope (KAIT), 
\citet{fcs+03} noted that there was a subclass of peculiar Type I supernovae
that appeared to be rich in Calcium at an early stage. \citet{pgm+10} presented
extensive observations of one of these events, SN\,2005E. 
%Rare. (We present available data on the remaining six candidates from LOSS in \S\ref{sec:loss}.) 
In addition to the Calcium-richness, SN\,2005E was subluminous and rapidly evolving. Furthermore, 
SN\,2005E was located 23\,kpc away from the nucleus of its host galaxy, with no signs
of underlying star formation to deep limits. The physical origin of SN\,2005E
remains a matter of debate.

Motivated thus, and given that the survey design of PTF allows it be to especially
sensitive to find and follow-up such events, we searched the PTF discovery stream.
We find three transients -- PTF\,09dav, PTF\,10iuv and PTF\,11bij -- that share five 
distinguishing characteristics with SN\,2005E: (i) peak luminosity intermediate 
between that of novae and supernovae, (ii) faster photometric evolution (rise and decline)
than supernovae, (iii) photospheric velocities comparable to supernovae,
(iv) early evolution to the nebular phase, and (v) nebular spectra dominated by
Calcium emission. These five explosion properties define a distinct class of 
``Calcium-rich gap'' transients. (We closely scrutinize the choice of this 
set of explosion characteristics in \S7.1.)

Remarkably, these three PTF transients were also found in remote locations, more than 30\,kpc
away from the nuclei of their host galaxies. Location was not used as a selection criterion; yet, 
we do not see any other PTF supernovae thus far that display the five explosion 
characteristics listed above (see \S7.2 for details). Therefore, in \S2 we take a 
closer look at the locations of these three transients in the context of locations 
of other PTF supernovae. 

The rest of the paper is organized as follows.  We present observations of 
PTF\,09dav, PTF\,10iuv and PTF\,11bij in \S\ref{sec:09dav}, \S\ref{sec:10iuv} 
and \S\ref{sec:11bij} respectively. We show the available data on 
the six archival candidates from the past decade in \S\ref{sec:loss}. 
We analyze the properties of the combined sample of well-observed 
transients in this class in \S\ref{sec:analysis}. We discuss whether this
class has an underlying physical commonality and speculate on its origin 
in \S\ref{sec:discussion}, and conclude in \S\ref{sec:conclusion}. 

\begin{figure*}[!hbt] 
   \centering
   \includegraphics[width=3.5in]{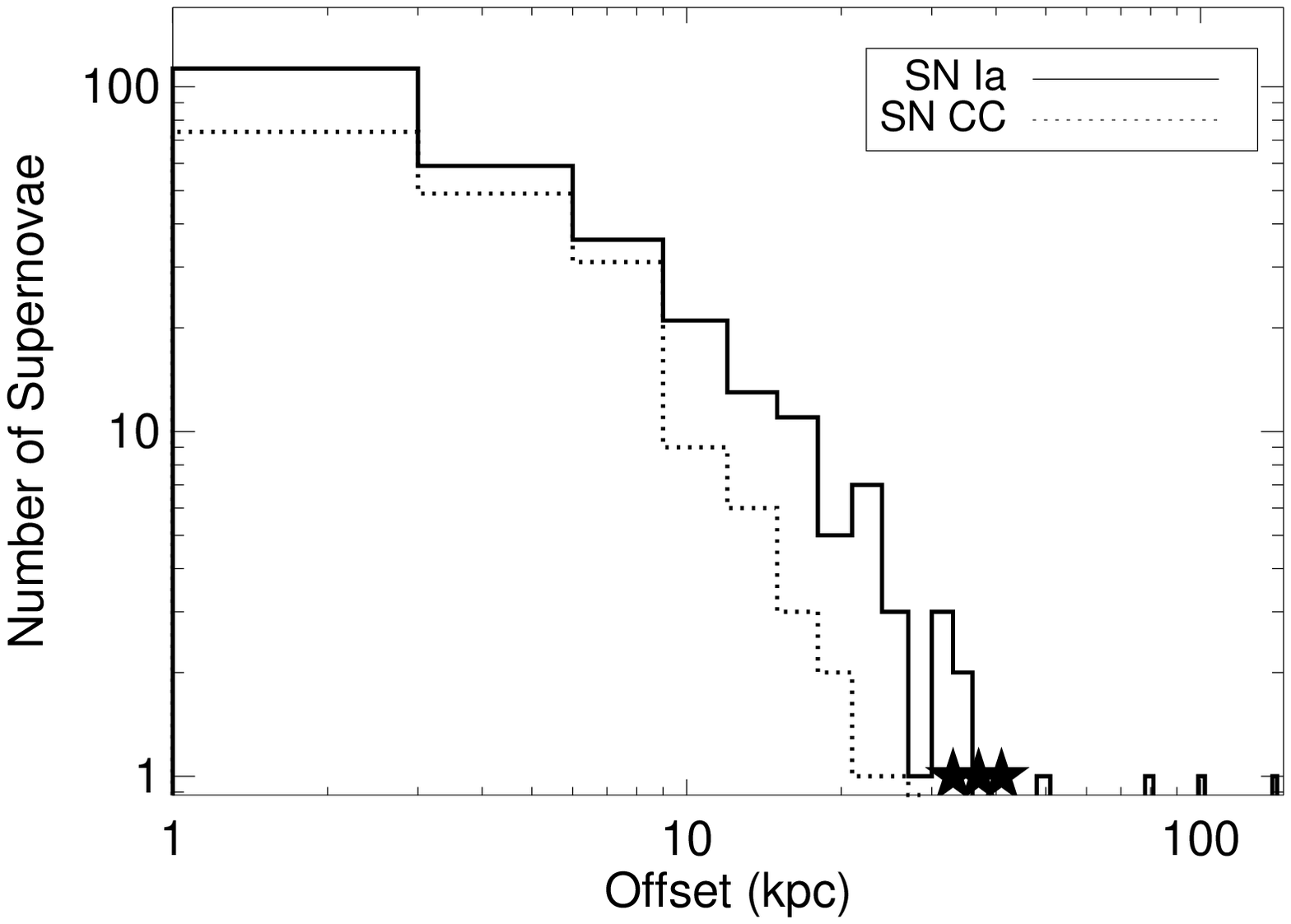}\includegraphics[width=3.5in]{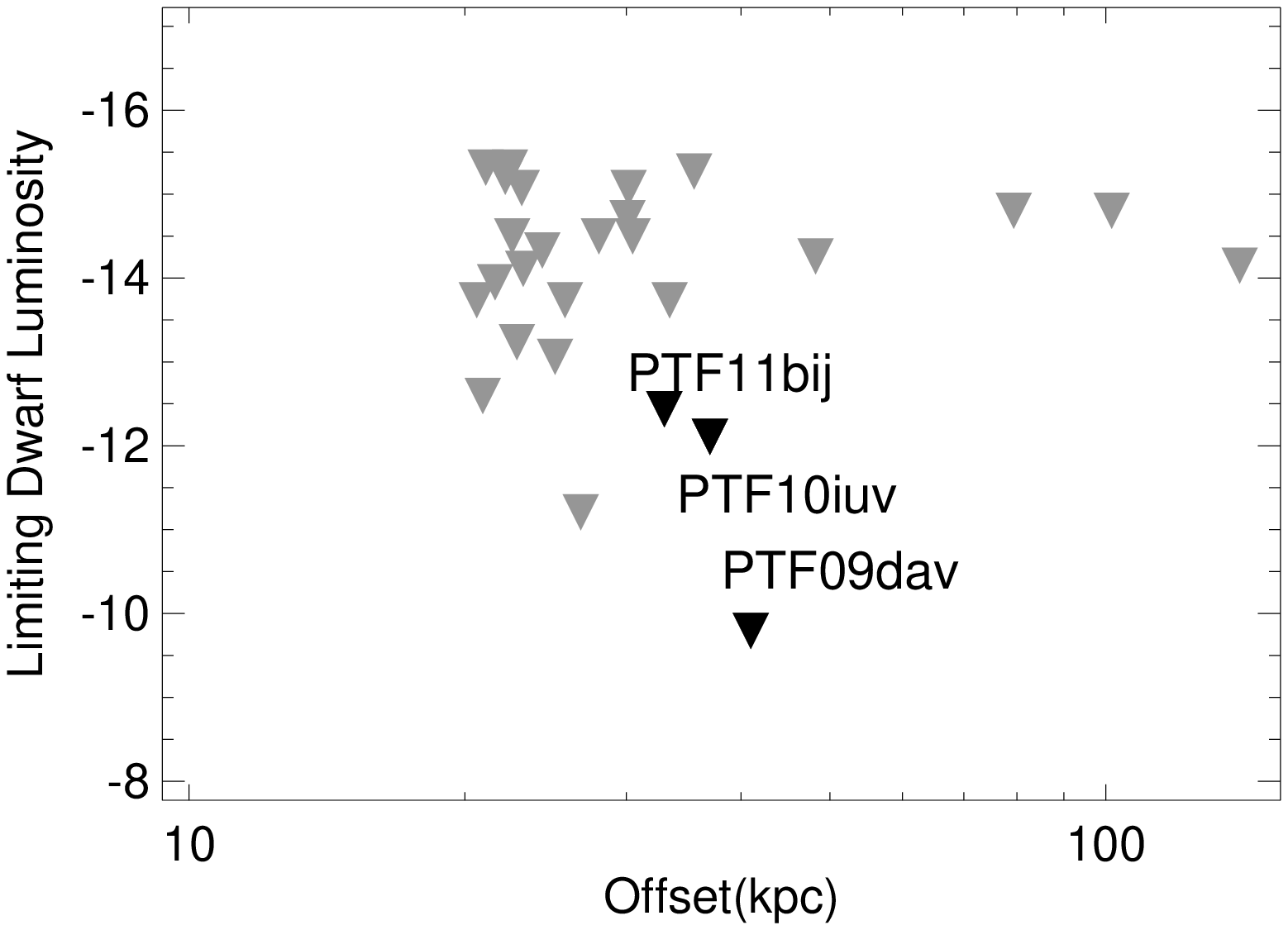}
   \caption[Offset Distribution of PTF Supernovae]{\small {\it Left:} Histogram of projected offsets from the host-galaxy nucleus for 
the first 520 transients discovered and spectroscopically classified by the Palomar Transient Factory with $z\,<\,$0.1. 
Stars denote the locations of PTF\,11bij, PTF\,10iuv and PTF\,09dav. 
%[{\bf add here typical galaxy light vs. kpc plots; convert to half-light radii where available}].
{\it Right:} Upper limits for an underlying dwarf host ($M_{R}$) for transients with offset greater than 20\,kpc.
}
\label{fig:snoffsets}
\end{figure*}

\section{Offset Distribution of PTF Supernovae}
\label{sec:offsets}
The location of a transient explosion has long been exploited as a clue
to determining its nature. It has been suggested that the classical nova population
is bimodal, with some dependence on whether the nova is located in the Galactic disk or bulge \citep{sdh+11}. 
Several studies of supernova host-galaxy properties as well as the site 
of the supernova  within the host galaxy have been undertaken 
\citep{h01,v97,psb+08,hmp+09,bp09,aj09}.
Subluminous Type Ia supernovae (SN\,Ia) are found preferentially in old environments \citep{h01}
(for a review of spectroscopic classification of supernovae, see \citealt{f97}).
Core-collapse supernovae (SN\,CC) and more luminous SN\,Ia are 
preferentially found in late-type galaxies \citep{hps+96}. Type Ic supernovae are not found in dwarfs 
\citep{agk+10} and SN\,Ibc are more centrally concentrated than SN\,II \citep{aj09}. 
About 15\% of the stellar mass is expected to be in the intergalactic medium in clusters, and a 
handful of intracluster supernovae have been discovered \citep{gmg+03,sgb+11}. At higher energies, gamma-ray burst
offsets from host galaxies offered a clue to their progenitors \citep{fls+06,bpp+06}.

In the first two years, PTF
discovered and spectroscopically confirmed over 1300 extragalactic transients. Here, we limit the study 
of the offset distribution of PTF supernovae to the first 520 transients which were nearby, with redshift 
$z < $0.1. We compute precise projected offsets from the host galaxies for each transient, and
the resulting histogram is shown in Figure~\ref{fig:snoffsets}. The redshift of the host galaxy is
consistent with the redshift of the transient. Note that the offset is projected and hence, only a
lower limit.  

We split the population into core-collapse supernovae and thermonuclear supernovae. 
We find that the core-collapse and thermonuclear populations 
show similar distributions out to a projected offset from the host nucleus 
of $\approx$\,9\,kpc suggesting that the average distribution is 
proportional to starlight. Beyond 9\,kpc, the thermonuclear population shows a heavily extended 
tail suggesting a second parameter governing their rate or perhaps even two progenitor populations
\citep{mdp06}. A cautionary note here is that although this sample is relatively more homogeneous in that it is drawn
from a single survey with extensive follow-up observations, it is not spared the vagaries of 
incompleteness due to weather and follow-up bias.
%(The details of the offset distribution for different galaxy morphologies and supernova
%sub-types is the subject of another paper).

Next, we take a closer look at the population with offsets larger than 20\,kpc.
We co-add available pre-explosion data to derive limiting magnitudes on a  
satellite host galaxy at the location of the supernova (see Figure~\ref{fig:snoffsets}, right panel).
The redshift limit constrains the luminosity of underlying host galaxies based on our pre-explosion 
co-adds to at least $-$16\,mag. 

Focusing on the dozen transients with offsets larger than 30\,kpc, we find
9 SN\,Ia, 0 SN\,CC, and 3 peculiar transients. Five SN\,Ia -- PTF\,09cex, 
PTF\,10fjg, PTF\,10xua, PTF\,10xgb, and PTF\,10xgc -- with offsets of 
30--36\,kpc have spectroscopically confirmed host redshifts. Three SN\,Ia
-- PTF\,10qnk (101\,kpc), PTF\,10qht (79\,kpc), and PTF\,10qyx (48\,kpc) -- are 
possibly intracluster supernovae (Horesh et al., in prep.). 
The furthest offset SN\,Ia, PTF10ops (140\,kpc), has peculiar properties
and is the subject of another paper \citep{mst+11}.  
%%10qyx (48kpc -- could this be intracluster, VVDS survey) 
%%10qht is intracluster (79 kpc)
%%{\bf A mask of PTF\,10qnk (101\,kpc) is pending once it rises}. 

The three peculiar transients are the topic of this paper.
PTF\,09dav is offset from a spiral host by 40\,kpc; 
PTF\,10iuv is in a galaxy cluster containing early and late-type galaxies, 
with the closest being an elliptical galaxy 37\,kpc away; and
PTF\,11bij is in a cluster of early-type galaxies, with the closest 
being 33\,kpc away. Note that in units of host-galaxy Petrosian radii,
these offsets correspond to 4.9, 5.9, and 3.8 radii, respectively.
We cannot rule out the possibility of an underlying low-luminosity dwarf host; we can only constrain dwarfs to a limiting $R$-band absolute magnitude of
$-$9.8, $-$12.1, and $-$12.4 for PTF\,09dav, PTF\,10iuv, and PTF\,11bij, respectively. 
Finding dwarf satellites at a separation of 40\,kpc would not be unusual 
(cf., Andromeda's $-$8.5\,mag satellite at 350\,kpc; \citealt{sbm11}). 

%We emphasize here that remote locations are not a selection criterion for identifying
%members of this class. As discussed in detail in \S7.2, there are no other PTF 
%transients that match the explosion properties of this class found in the disks of galaxies.

%%% Petrosian Radii
%PTF11bij 12.665'', 1.44''/kpc => 3.8 radii
%PTF10iuv 13.390'', 18.362, 2.15''/kpc => 5.9 radii
%PTF09dav 11.312'', 1.39''/kpc => 4.9 radii
%SN2005E 19.304''

\section{Observations: PTF\,09dav}
\label{sec:09dav}
We presented the discovery, light curve, and photospheric 
spectra of PTF\,09dav in a previous paper \citep{skn+11}.
We found that PTF\,09dav had a low peak absolute magnitude of $M_{R}\,\approx\,-$16.3
and a short rise-time of 12\,days relative to SN\,Ia (which peak at $-$19\,mag with
a rise-time of 17.5\,days). We also found that the photospheric 
spectra resembled those of SN\,1991bg-like \citep{frb+92} subluminous SN\,Ia but with very unusual strong lines of
\ion{Sc}{2}, \ion{Mg}{1} and possibly \ion{Sr}{2}. Photospheric velocities were $\approx\,$6000
km\,s$^{-1}$.
Here, we present late-time imaging and nebular spectroscopy of PTF09dav.

\begin{figure*}[!hbt]
\begin{center}
\includegraphics[width=7.5in]{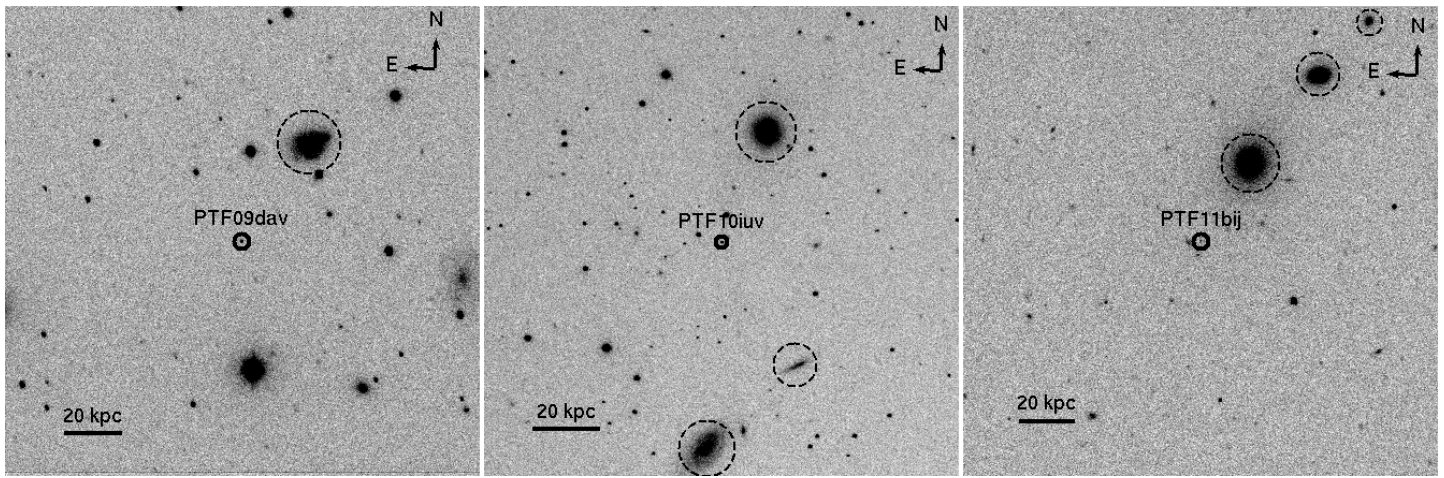}
\begin{scriptsize}
\caption[Outskirts location of PTF\,09dav, PTF\,10iuv and PTF\,11bij]{\small {\it Left:} 
PTF\,09dav is offset from its late-type host (circled) by 40\,kpc. Note that the galaxy
on the western edge is unrelated as it is at a different redshift. 
 {\it Center:} PTF\,10iuv is offset from a galaxy group (circled) with early-type
and late-type galaxies; the nearest potential host is 37\,kpc away. 
{\it Right:} PTF\,11bij is offset from a group of early-type galaxies (circled);
the nearest is 33\,kpc away.
}
\end{scriptsize}
\end{center}
\label{fig:loc}
\end{figure*}

\subsection{Late-Time Imaging}
\begin{figure}[htbp] 
   \centering
   \includegraphics[height=2.7in]{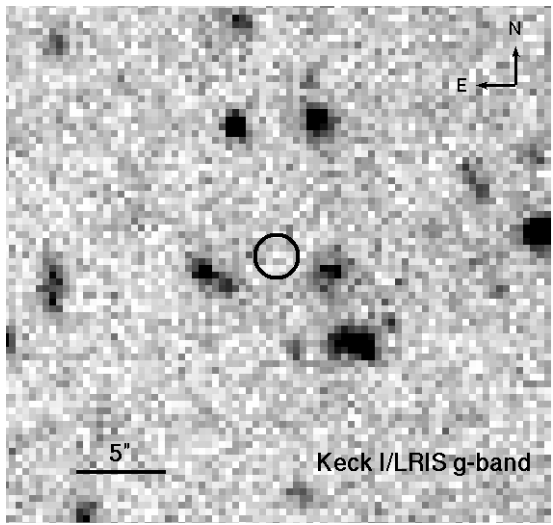}
   \caption[Deep limit on dwarf host for PTF\,09dav]{\small Deep late-time $g$-band imaging 
with Keck I/LRIS  showing no host galaxy 
at the position of PTF\,09dav brighter than 26\,mag. The registration accuracy is %0.165$\arcsec$ in $R$-band 
0.241$\arcsec$. We denote the position of PTF\,09dav with a 5$\sigma$ position error circle.
}
\label{fig:gR09dav}
\end{figure}

We undertook deep imaging in the $g$-band and $R$-band filters at the 
position of PTF\,09dav with the Low Resolution Imaging Spectrometer
(LRIS; \citealt{occ+95}) on the Keck I telescope on 2010 May 15.603 
and July 9.584 (UT dates are used throughout this paper), 9--11 months after explosion. 
%under seeing conditions of xx arcsec and xx arcsec respectively. 
We registered these images with a Palomar 60-inch (P60) image of PTF\,09dav. 
No source is detected to a  3$\sigma$ limiting magnitude of $R = 26.2$\,mag 
(Table~\ref{tab:lrislimits}, Figure~\ref{fig:gR09dav}). This constrains 
any satellite, dwarf host to be fainter than $M_R = -9.8$\,mag.

We undertook imaging in the $K^{\prime}$ band with Laser Guide Star Adaptive Optics 
(LGS-AO; \citealt{wcj+06,vbl+06}) on the Keck II telescope with the Near Infrared Camera 2 
(NIRC2). On 2010 June 17.568, we obtained 10 images of 10\,s co-added integrations.
The zeropoint was derived relative to the 2MASS catalog \citep{scs+06}. No source was detected 
to a 3$\sigma$ limit of $K^{\prime} = 21.1$\,mag.

\begin{deluxetable*}{llllll}%[!hbt]
  \tabletypesize{\footnotesize}
  \tablecaption{Late-Time Photometry of PTF\,09dav}
  \tablecolumns{6}
  \tablewidth{0pc}
 \tablehead{\colhead{Date} & \colhead{Phase} & \colhead{Facility} & \colhead{Exposure} & \colhead{Filter} & \colhead{Magnitude}\\
\colhead{(UT 2010)} & \colhead{(day)} & \colhead{} & \colhead{(s)} & \colhead{} & \colhead{} 
}
 \startdata
May 15.612 & 279.7 & Keck I/LRIS   & 1230 & $g$ & $>$26.0 \\ %30s, 300s*4 
May 15.614 & 279.7 & Keck I/LRIS   & 1030 & $R$ & $>$24.8 \\ %30s, 250s*4 
June 17.568 & 313.7 & Keck II/NIRC2 &  100  & $K^{\prime}$ & $>$21.1           \\ %55365.5677, 10s x 10
July 9.584 & 334.7 & Keck I/LRIS   & 1450 & $g$ & $>$25.4  \\  %290s*5 exposures
July 9.584 & 334.7 & Keck I/LRIS   & 1200 & $R$ & $>$26.2           %240s*5 exposures   
 \enddata
\label{tab:lrislimits}
\end{deluxetable*}
%%Epoch of maximum light in B-band is 2009 Aug 8.9 or MJD 55051.9

\subsection{Nebular Spectroscopy}
On 2009 November 11, only three months after maximum light, a spectrum
taken with LRIS on Keck I revealed that PTF\,09dav had become nebular. The 
time scale to become nebular was surprising as it was faster 
than that of typical supernovae by a factor of a few. 
Furthermore, only two emission features are seen -- H$\alpha$ and [Ca~II] $\lambda\lambda$\,7291, 7324). The width, 
redshift, and flux of the lines are summarized in Table~\ref{tab:nebular}.
%H$alpha$ is
%seen at 6820.36\AA\ with a width of 29.07\AA\ and a flux of 7.62$\times$10$^{-17}$ (equivalent width of $-$143.6). 
%The [Ca II] line is seen at 7579.07\AA\ 
%with a width of 167.1\AA\ and a flux of 2.03$\times$10$^{-15}$ (equivalent width of $-$5132). Corrected for instrumental 
%resolution, widths are 23.5\AA\ and 166.1\AA\ respectively. The width and redshift of H$\alpha$ is 1000 km\,s$^{-1}$ and 
%843 km\,s$^{-1}$. The width and redshift
%of [Ca II] is 6600 km\,s$^{-1}$ and 250 km\,s$^{-1}$ (assuming z=0.0363 and averaging the doublet wavelengths of 7291\AA\ and 7324\AA\ ). 
The absence of the Ca~II near-infrared (IR) triplet ($\lambda\lambda$\,8498, 8542, 8662) is indicative
of a low circumstellar density. No H$\alpha$ was seen in the photospheric spectra presented
by \citet{skn+11}. The presence of H$\alpha$ emission is usually
interpreted as the interaction of a massive star wind with the circumstellar
environment.  Neither oxygen (prominent in SN\,CC) nor
iron lines (prominent in SN\,Ia) are seen in the nebular phase. 
%However, the location of the transient so far from its
%putative host makes the massive star origin unlikely.
These characteristics of the nebular spectrum were unprecedented
and no match could be found in supernova libraries.

\begin{figure*}[htbp] 
   \centering
   \includegraphics[width=5in]{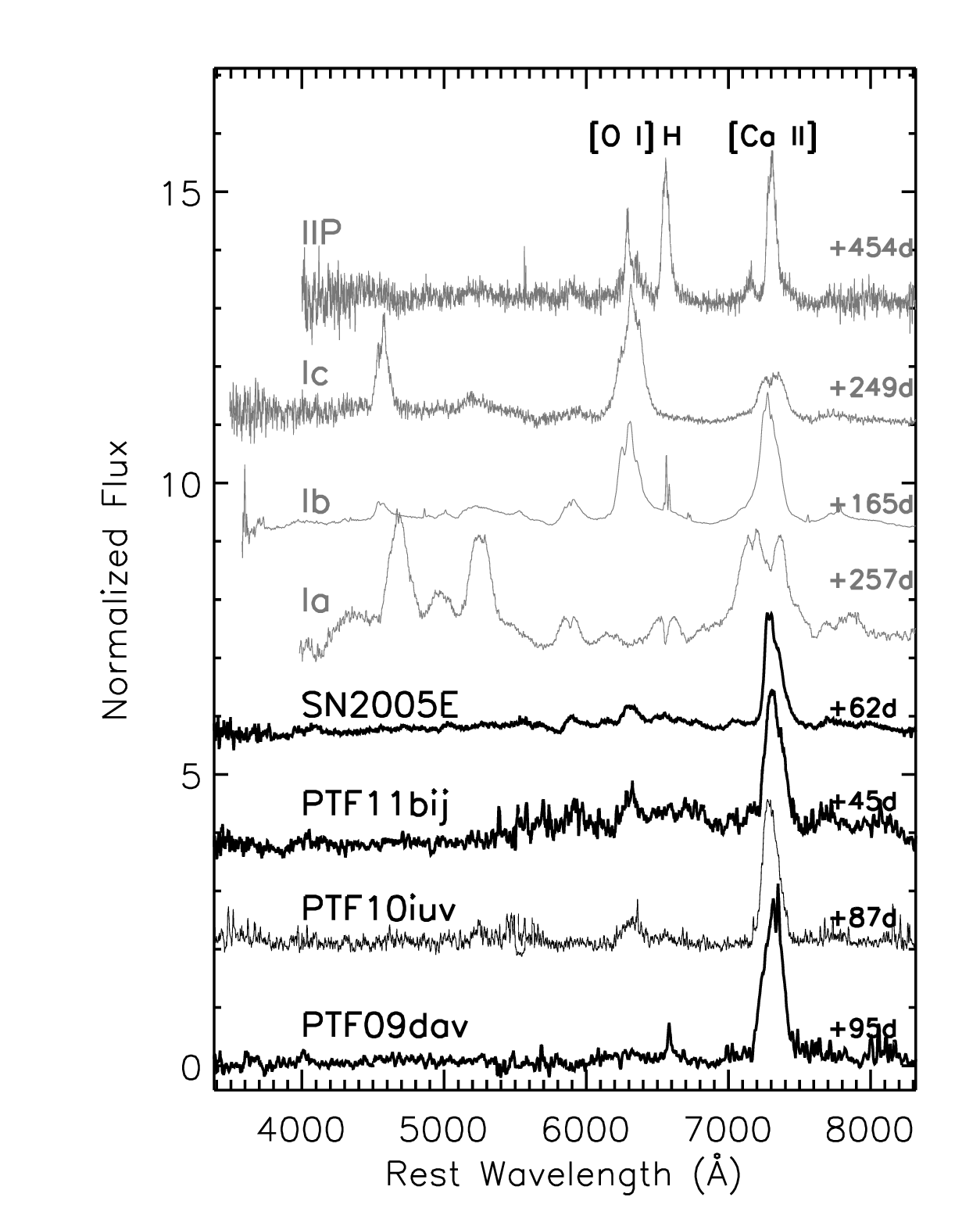} 
   \caption[Peculiar Nebular Spectra of PTF\,09dav, PTF\,10iuv, and SN\,2005E]{\small Nebular spectra of PTF\,09dav, PTF\,10iuv, and SN\,2005E. All three are very Calcium-rich.
Also shown for comparison, nebular spectra of a SN\,Ia (SN\,1986G, +257\,day; Asiago Catalog), 
a SN\,Ib (SN\,2007C, +165\,day; \citealt{tvb+09}), a SN\,Ic (SN\,2002ap, +249\,day; \citealt{gos+02}) and 
a SN\,IIP (SN\,2004et, +454\,day; \citealt{sas+06}). The nebular spectra of SN\,Ia
are dominated by [Fe~II], [Fe~III], and  [Co~III] emission lines. The nebular spectra of
SN\,CC have a much lower ratio of Calcium to Oxygen. 
}
\label{fig:nebspec}
\end{figure*}

%%Notes: http://adsabs.harvard.edu/abs/2010ApJ...708.1703M has 12 Type Ia nebular modelled, 
%http://adsabs.harvard.edu/abs/2009MNRAS.397..677T has 39 Type Ib/c nebular spectra
%Table 4 of http://adsabs.harvard.edu/abs/1996MNRAS.283....1T has a line list for [Fe] and [Co]
%http://adsabs.harvard.edu/abs/1997MNRAS.284..151M models and labels spectra at all phases of Type Ia

\section{Observations: PTF\,10iuv}
\label{sec:10iuv}
\subsection{Discovery and Light Curve}
On 2010 May 31.241, PTF discovered a new transient,
PTF\,10iuv, at $\alpha$(J2000) = 17$^h$16$^m$54.27$^s$ and
$\delta$(J2000) = $+31^\circ 33' 51.7''$.
It was initially found at $R = 21.2$\,mag, and we monitored its brightness
with the P60 in the $Bgriz$ filters for three
months \citep{cfm+06}. Late-time photometric observations were taken with the Large Format Camera
(LFC) on the Palomar 200-inch telescope and LRIS on the Keck I telescope. 

Data were reduced following standard procedures and aperture photometry was performed. 
Photometric calibration was done relative to photometry of field stars from the 
Sloan Digital Sky Survey \citep{aaa+09}. A common set of calibration stars was chosen for the 
P48, P60, LFC and LRIS data for consistency. Conversion from $ugriz$ to the $B$ band was done 
following \citet{jga06}. 

\begin{figure}[!htbp] 
   \centering
   \includegraphics[width=3.0in]{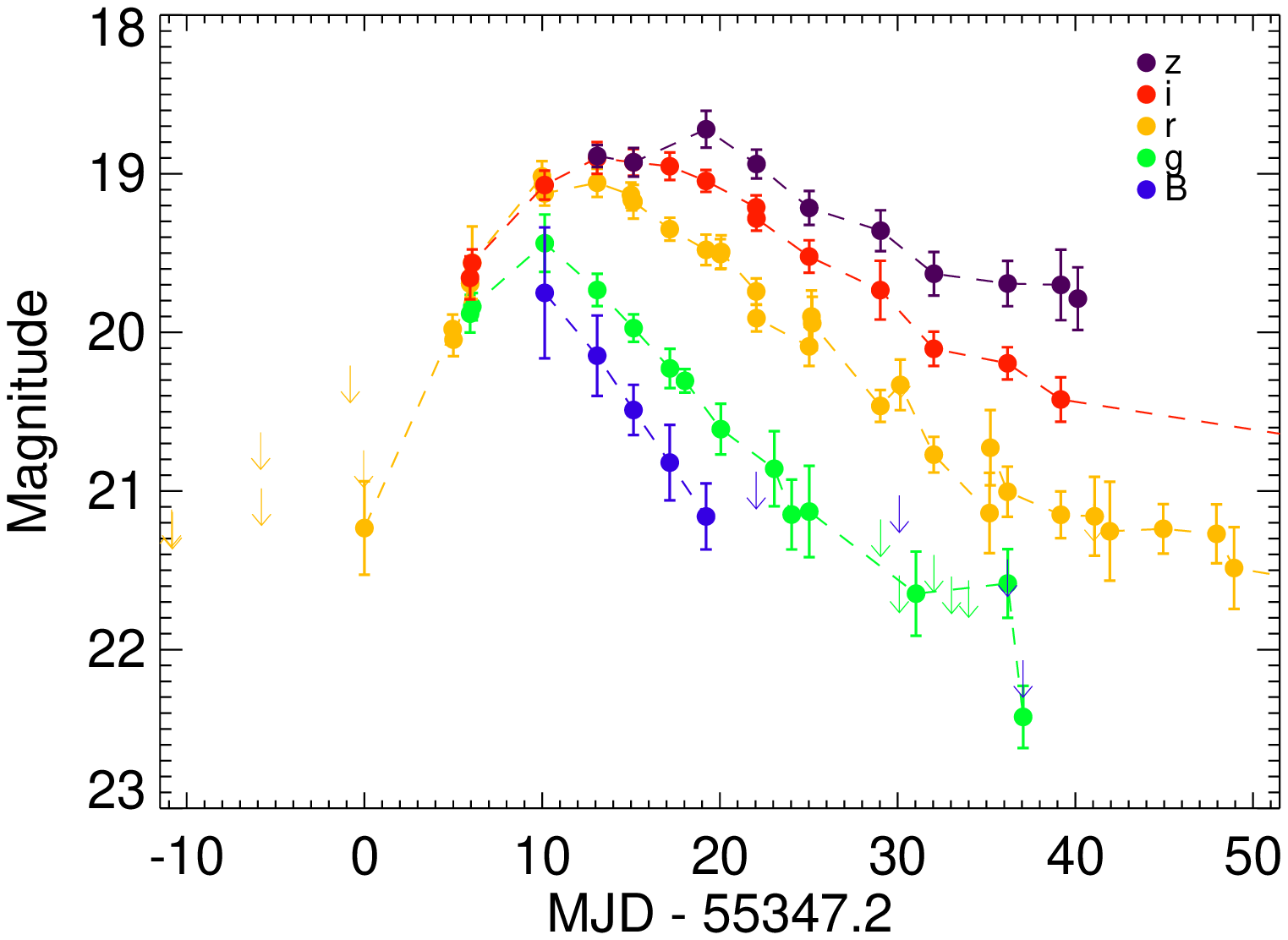} 
   \caption[Light Curve of PTF\,10iuv]{\small Light curves of PTF\,10iuv. Note the rapid rise of 3\,mag in 12\,days,
followed by the rapid decline at the rate of 1\,mag in 12\,days. 
}
\label{fig:10iuvlc}
\end{figure}

The light curve is summarized in Figure~\ref{fig:10iuvlc}. % and Table~\ref{tab:10iuvlc}.
PTF\,10iuv peaked on June 10 with $R = 19.0$\,mag. It rapidly rose by 3\,mag in 12\,days,
followed by a rapid decline at the rate of 1\,mag in 12\,days for one month. Subsequently, PTF\,10iuv
evolved slowly at the rate of 0.02 mag day$^{-1}$ for three months followed by 0.005 mag day$^{-1}$.
The color was neither extremely red nor blue. It evolved from $g-i \approx 0.4$ near maximum 
to $g-i \approx 0.7$ one month later.

\subsection{Spectroscopy}
\begin{figure}[!htbp] 
   \centering
   \includegraphics[width=3.5in]{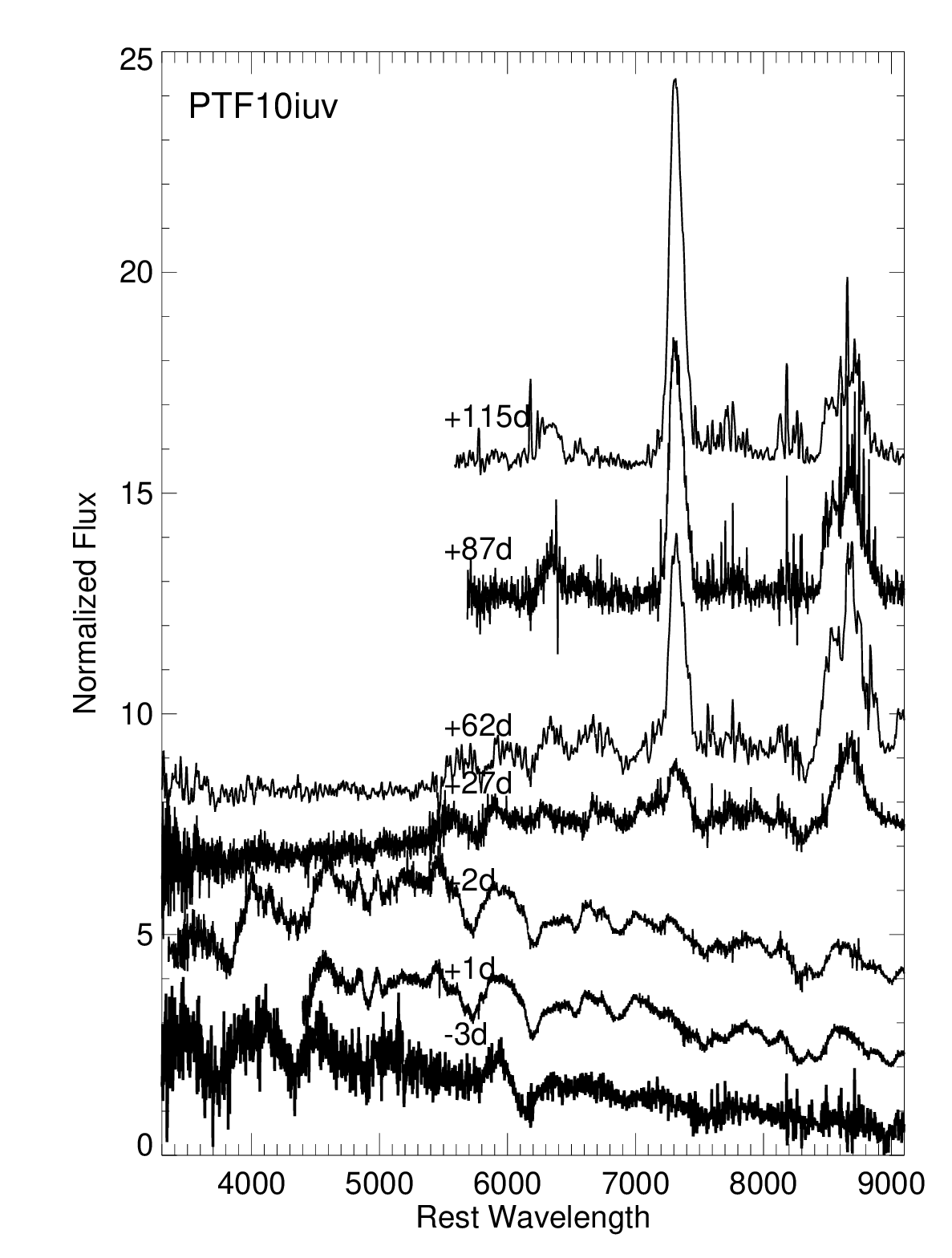}  
   \caption[Spectral Evolution of PTF10iuv]{\small Spectral evolution of PTF\,10iuv.
}
\label{fig:spec10iuv}
\end{figure}

On June 7, we obtained a classification spectrum using ISIS on the William Herschel  Telescope (WHT). Subsequently,
we continued to monitor the evolution with the Keck I (+LRIS) and Keck II (+DEIMOS) telescopes
until the spectra became completely nebular (Figure~\ref{fig:spec10iuv}).

\begin{figure}[!htbp] 
   \centering
   \includegraphics[width=3.0in]{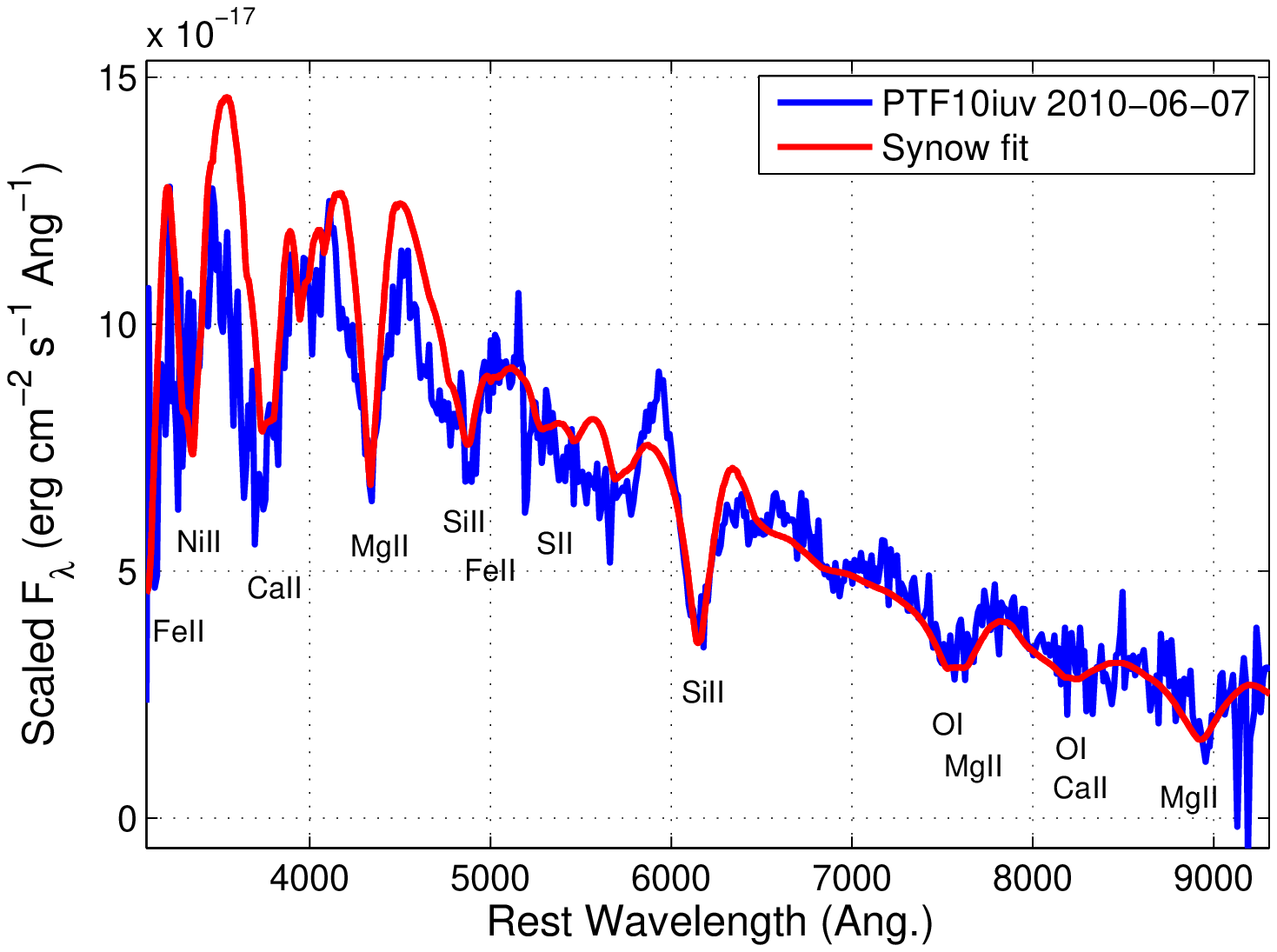} \\
   \includegraphics[width=3.0in]{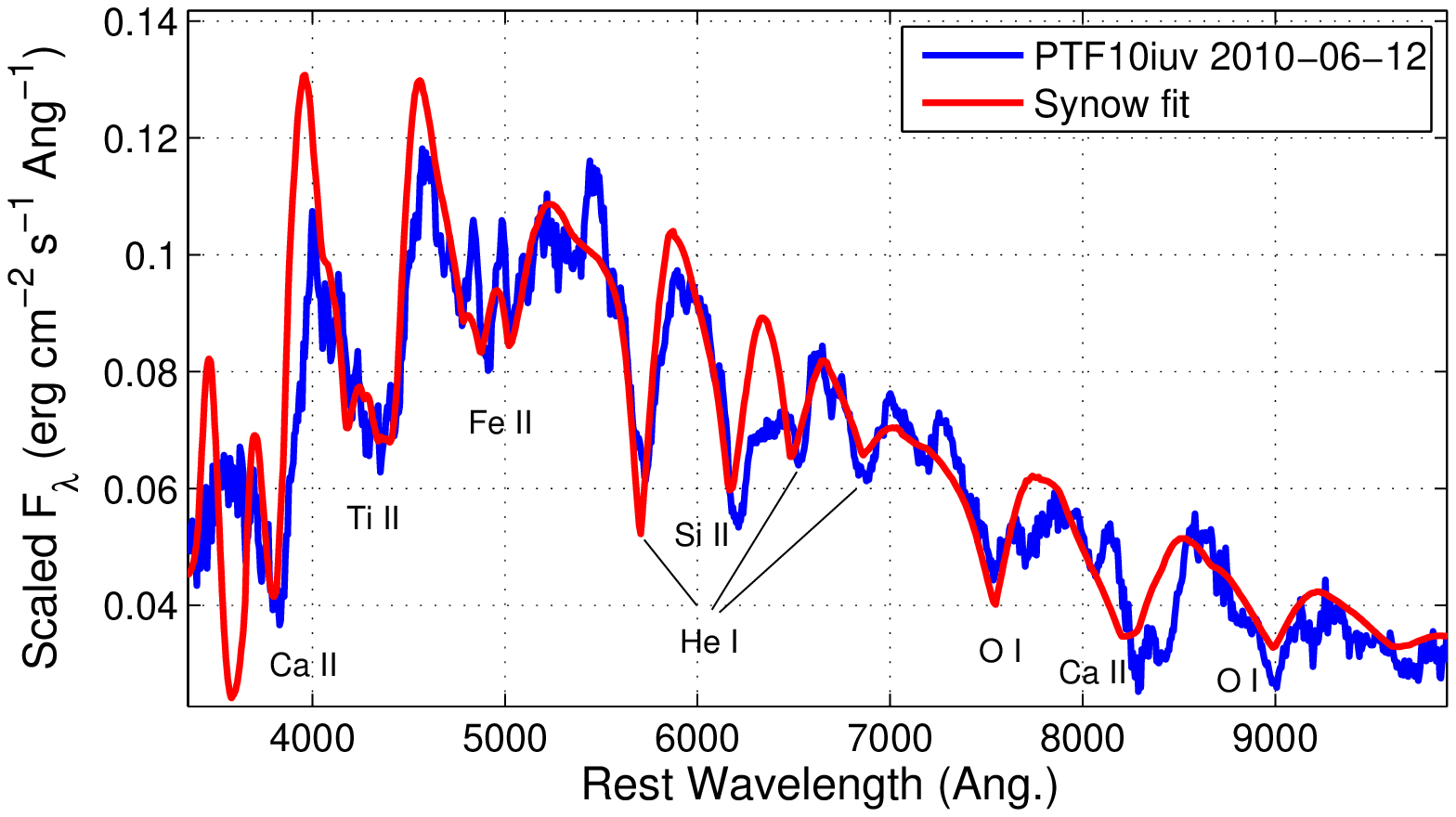}
   \caption[SYNOW fits to PTF10iuv]{\small SYNOW fits to photospheric spectra of PTF\,10iuv. 
{\it Top:} SYNOW fit of a pre-maximum spectrum ($T\,\approx\,$8500\,K, $v\,\approx\,$10,000\,km\,s$^{-1}$) with
prominent Magnesium, Silicon and Oxygen lines. 
{\it Bottom:} SYNOW fit of post-maximum photospheric spectrum of PTF\,10iuv ($T\,\approx\,$6500\,K, $v\,\approx\,$9000\,km\,s$^{-1}$).
The most prominent features are Helium, Oxygen, Calcium and Titanium.
}
\label{fig:synow}
\end{figure}

The spectra evolved to show prominent Helium features, resembling Type Ib spectra. The Calcium
lines become stronger with time, especially [Ca~II] relative to O~I. We identify major lines
both pre-maximum and post-maximum using SYNOW (Figure~\ref{fig:synow}).

As with PTF\,09dav, the spectrum of PTF\,10iuv became nebular in only 3 months. Both showed
prominent Calcium emission (Table~\ref{tab:nebular}, Figure~\ref{fig:nebspec}). Unlike
PTF\,09dav, PTF10\,iuv showed O~I emission and did not exhibit H$\alpha$ emission in the nebular phase.

\subsection{Radio Observations}
We observed PTF\,10iuv with the Expanded Very Large Array twice, once on
2010 August 25.06 and again on 2011 May 12.20. We observed in the X-band (8.46\,GHz) 
and added together two adjacent 128\,MHz subbands with full polarization to maximize continuum
sensitivity. Amplitude and bandpass calibration was achieved using
the archival flux value of J1721+3542. Phase calibration was carried
out every 10\,min by switching between the target field and the
point source J1721+3542. The visibility data were calibrated and imaged
in the {\it AIPS} package following standard practice. 

The transient was not detected with a 3$\sigma$ upper limit of 189\,$\mu$Jy during
the first epoch and 96\,$\mu$Jy during the second epoch.
This corresponds to $L_{\nu} < 1.0 \times 10^{27}$\,erg\,s$^{-1}$\,Hz$^{-1}$.
%Note from Dale: a separate flux calibrator was not observed as during this phase 
%of EVLA commissioning, a large overhead of 10-12min was needed to observe calibrators.

\begin{deluxetable*}{lllllll}%[!hbt]
  \tabletypesize{\footnotesize}
  \tablecaption{Lines in Nebular Spectra}
  \tablecolumns{7}
  \tablewidth{0pc}
 \tablehead{\colhead{Transient} & \colhead{Ion} & \colhead{Line Center} & \colhead{Shift$^{a}$} & \colhead{Flux} & \colhead{Width$^{b}$} & \colhead{Velocity Width} \\
\colhead{} & \colhead{} & \colhead{\AA\ } & \colhead{km\,s$^{-1}$} & \colhead{erg\,cm$^{-2}$\,s$^{-1}$} & \colhead{\AA\ } & \colhead{km\,s$^{-1}$} 
}
 \startdata
SN\,2005E  & O~I       & 6360.6 & $-$1120 & 2.0$\times$10$^{-15}$ & 107.8  & 5110  \\
SN\,2005E  & [Ca~II]   & 7367.0 & $-$17   & 1.5$\times$10$^{-14}$ & 127.6  & 5240  \\ %34Mpc, z=0.0082, DM=32.7
SN\,2005E  & Ca~II     & 8710.4 & 2530    & 2.7$\times$10$^{-14}$ & 299.0  & 10470 \\
PTF\,09dav & H$\alpha$ & 6820.4 & 840     & 7.6$\times$10$^{-17}$ & 23.5   & 1000  \\
PTF\,09dav & [Ca~II]   & 7579.1 & 250     & 2.0$\times$10$^{-15}$ & 166.1  & 6600  \\
PTF\,10iuv & O~I       & 6467.7 & $-$460  & 1.7$\times$10$^{-16}$ & 170.8  & 8090  \\
PTF\,10iuv & [Ca~II]   & 7465.4 & $-$410  & 8.0$\times$10$^{-16}$ & 134.2  & 5510  \\ %z=0.023
PTF\,10iuv & Ca~II     & 8827.3 & 2150    & 9.4$\times$10$^{-16}$ & 312.8  & 10950 \\ 
PTF\,11bij & O~I       & 6531.5 & $-$1010 & 8.6$\times$10$^{-17}$ & 98.8   & 4680  \\ %z=0.035
PTF\,11bij & [Ca~II]   & 7573.5 & 406     & 5.8$\times$10$^{-16}$ & 136.9  & 5620  \\
PTF\,11bij & Ca~II     & 8960.0 & 3140    & 1.3$\times$10$^{-15}$ & 257.0  & 9000  
%Ca lines = [7291,7324,8498,8542,8662]
%[O I] 6300-6364
 \enddata
\tablecomments{$^{a}$ Shift is computed relative to the velocity of the putative host galaxy.}
\label{tab:nebular}
\end{deluxetable*}

\section{Observations: PTF\,11bij}
\label{sec:11bij}
\begin{figure*}[!htbp] 
   \centering
   \includegraphics[width=5in]{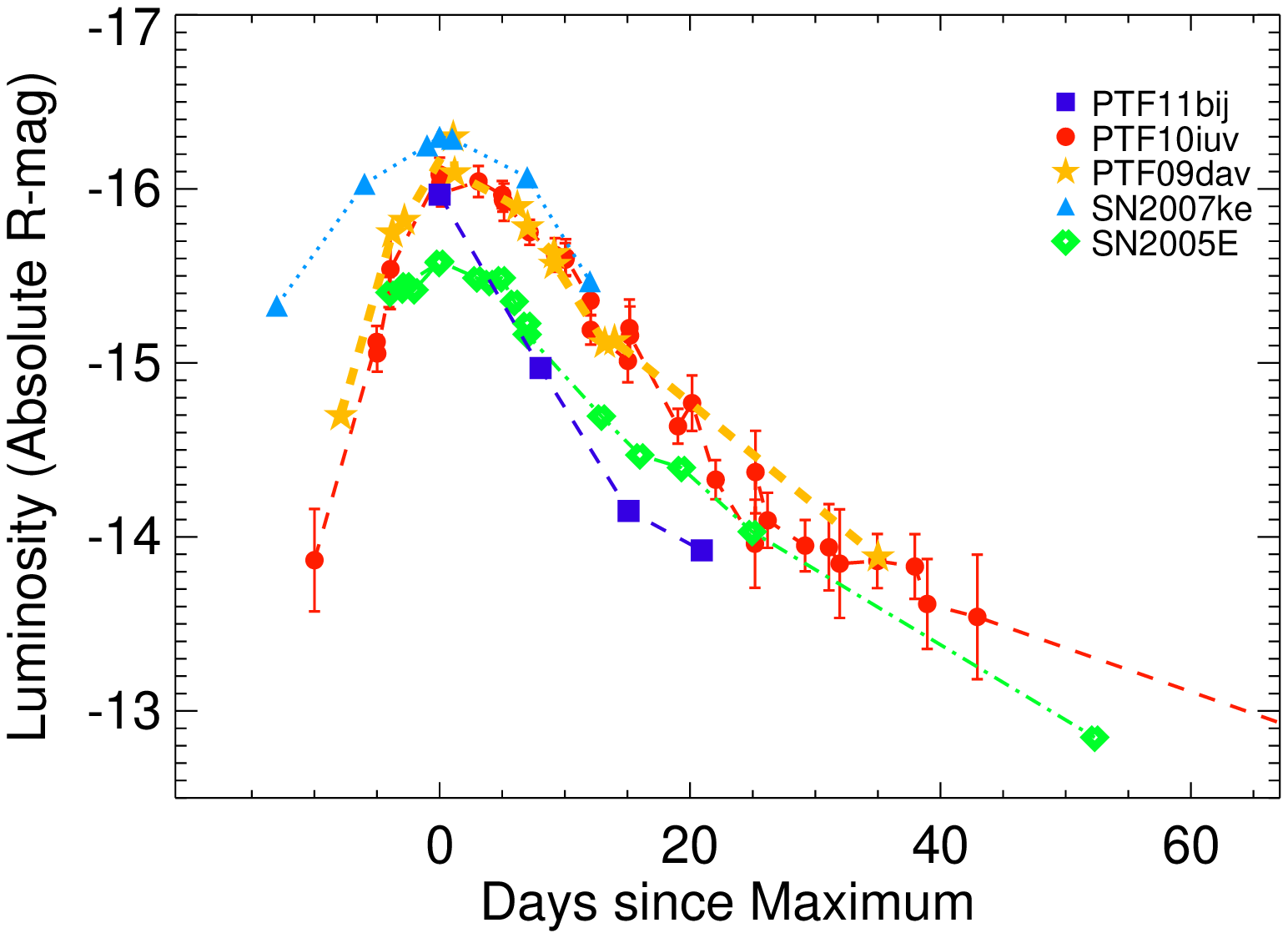}  
   \caption[Light Curves of Full Sample]{\small Light curves of all five members of the class of
Calcium-rich gap transients. Note the lower peak luminosities and faster evolution relative to normal
supernovae.
}
\label{fig:lcall}
\end{figure*}

%{\bf Need to check earlier images in March, especially March 9?}
On 2011 March 13.187, PTF discovered a new transient,
PTF\,11bij, at $\alpha$(J2000) = 12$^h$58$^m$58.39$^s$ and
$\delta$(J2000) = $+37^\circ 23' 12.0''$.
It was initially found at $R = 19.9$\,mag, and we monitored its brightness
with the P60 in the $gri$ filters.
Similar to PTF\,09dav and PTF\,10iuv, the light curve showed a peak luminosity
in the gap between novae and supernovae, and it evolved rapidly (see Figure~\ref{fig:lcall}). % and Table~\ref{tab:10iuvlc}.

We obtained a Target Of Opportunity spectrum
on 2011 April 3.40 and a nebular spectrum on April 27.40, 
both with  LRIS on the Keck I telescope (Figure~\ref{fig:nebspec}).
The spectra were Calcium-rich, similar to PTF\,10iuv and PTF\,09dav, 
and the redshift of 0.035 is consistent with a group of early-type galaxies
(the nearest galaxy is offset by 33\,kpc).

\begin{figure*}[!htbp] 
   \centering
   \includegraphics[width=5.5in]{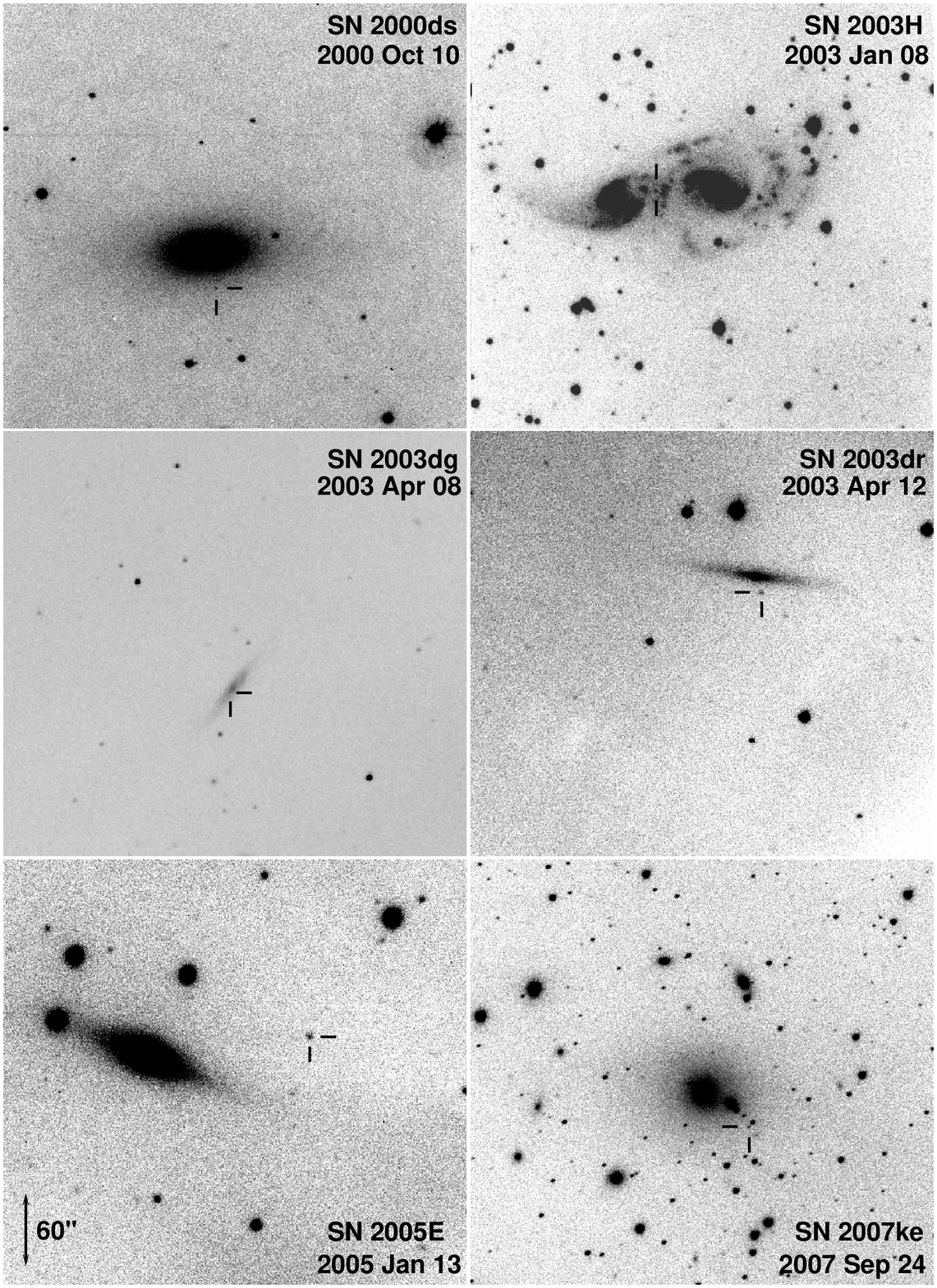} 
   \caption[Finder Charts for KAIT Sample]{\small Locations of Calcium-rich supernovae from the KAIT Lick Observatory
Supernova Search sample. From {\it Top Left} to {\it Bottom Right}: SN\,2000ds, SN\,2003H, SN\,2003dg, SN\,2003dr, SN\,2005E, and SN\,2007ke.
Unfortunately, adequate photometric data are available only for SN\,2005E and SN\,2007ke to confirm
their similarity to PTF\,09dav, PTF\,10iuv, and PTF\,11bij. Note that the locations are atypical: 
elliptical host, interacting galaxy pair environment, off the disk of an edge-on galaxy, 
large projected offset from isolated host, and galaxy group environment.
}
\label{fig:lossfinders}
\end{figure*}

\section{Observations: Archival Candidates}
\label{sec:loss}
Here, we present archival data on candidate transients from
other surveys noted for their Calcium-rich spectra
\citep{fcs+03}. Specifically, \citet{pgm+10} suggest that
the following six supernovae resemble SN\,2005E: 
SN\,2000ds (IAUC\#7507), SN\,2001co (IAUC\#7643), 
SN\,2003dg (IAUC\#8113), SN\,2003dr (IAUC\#8117), 
SN\,2003H (IAUC\#8045), and SN\,2007ke (CBET\#1084).  
SN\,2000ds was discovered by T. Puckett. SN\,2001co, SN\,2003dg, SN\,2003H, and SN\,2007ke were
discovered by LOSS. SN\,2003dr was discovered independently
by both T. Puckett and LOSS. 

\begin{figure}[!htbp] 
   \centering
   \includegraphics[width=3.5in]{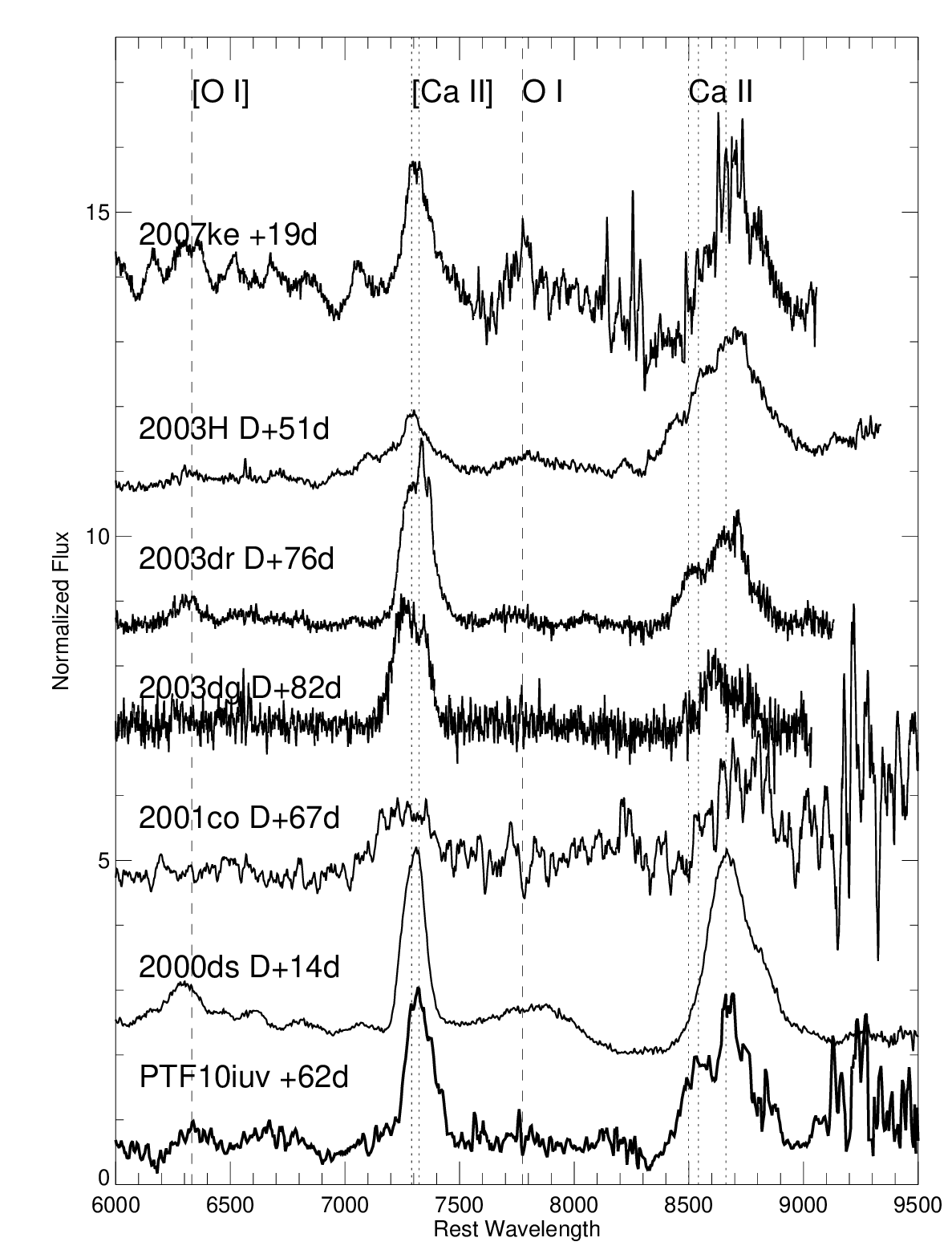}
   \caption[Spectra of KAIT Sample]{Spectra of the KAIT/LOSS sample of Calcium-rich supernovae compared to PTF\,10iuv. 
Except for SN\,2007ke, as the photometry is not available to constrain the maximum-light epoch, the phase is 
relative to discovery and not maximum. 
}
\label{fig:specloss}
\end{figure}

Here, we review the available data for these supernovae. In Figure~\ref{fig:lossfinders}, we display 
the environments of these candidate members of the Calcium-rich class. Next, in Figure~\ref{fig:specloss}, we plot their spectra 
(at the latest phase available) and compare them to those of PTF\,10iuv.
The strength of [Ca~II] relative to Oxygen is a distinguishing
spectroscopic feature of this class; it is common to SN\,2000ds, SN\,2003dg, SN\,2003dr, 
SN\,2007ke, and SN\,2003H. 
The case for SN\,2001co is less clear due to the noisy spectrum and we exclude it from further analysis.

Unfortunately, the photometry is extremely sparse for this sample --- % (Table~\ref{tab:lossphot}). 
only one of the five supernovae has more than two points on the light curve.
SN\,2007ke has a peak absolute magnitude of $-$16.3, a rise-time of 15\,days, and a decline 
rate of 0.1 mag day$^{-1}$ (Figure~\ref{fig:lcall}). Therefore, SN\,2007ke satisfies all of the
photometric and spectroscopic properties of the class of ``Calcium-rich gap"  transients
as outlined in \S1. Note that the location of SN\,2007ke is ``off-galaxy'' (Figure~\ref{fig:lossfinders},
bottom-right panel). In view of the paucity of photometric data for the other four supernovae,
we do not discuss them further.  

%The collective properties of PTF\,09dav, PTF\,10iuv and PTF\,11bij
%are consistent with only one other transient in the
%literature --- SN\,2005E \citep{pgm+10}. SN\,2005E
%was thus the first, well-studied member of this class 
%of ``Calcium-rich gap'' transients. It had a peak absolute
%magnitude in the gap (M$_{R}\approx\,-$15.5) and evolved rapidly.
%It had photospheric velocities of 11000 km\,s$^{-1}$ and calcium-rich nebular spectra.   

\section{Analysis}
\label{sec:analysis}

\subsection{The Sample}
To summarize, the five distinguishing characteristics of our 
peculiar new class of transients are (i) peak luminosity intermediate between that of
novae and most supernovae, (ii) faster photometric evolution (rise and decline)
than that of normal supernovae, (iii) photospheric velocities comparable to those of normal supernovae,
(iv) early evolution to the nebular phase, and (v) nebular spectra dominated by
Calcium emission.

Based on available observations, the objects that satisfy the defining
characteristics of this class of ``Calcium-rich gap'' transients are 
SN\,2005E, SN\,2007ke, PTF\,09dav, PTF\,10iuv, and
PTF\,11bij. Their light curves are shown in Figure~\ref{fig:lcall} and nebular
spectra in Figure~\ref{fig:nebspec}. Observed properties of these events are 
summarized in Table~\ref{tab:summary}.

Our choice of these five characteristics is motivated by choosing the maximum
set of characteristics that gives the largest number of related transients. 
Evidently, if we add characteristics to this list, the family size becomes smaller
and we may overlook a unifying explanation. If we remove characteristics from 
this list, the family becomes larger but we may inadvertently introduce a diversity 
of progenitor channels. We discuss examples of each case below.

If we require an additional sixth property of a photospheric spectrum with prominent Helium
lines, only PTF\,10iuv and SN\,2005E would be members of the family. It is not clear whether 
the photospheric spectra of PTF\,09dav exhibit Helium (see discussion in \citealt{skn+11}). The 
earliest spectra of SN\,2007ke and PTF\,11bij are not until +19\,d and +21\,d after maximum light. 
These two spectra already show the nebular features of prominent [\ion{Ca}{2}] emission common to 
the class, but we cannot confirm Helium. Therefore, we allow a photospheric diversity but require 
a nebular uniformity in the class.

If in addition to Calcium, we also require Oxygen emission in the nebular spectrum, only PTF\,10iuv, 
SN\,2005E, and PTF\,11bij would be members of this class. PTF\,09dav does not show Oxygen emission
and SN\,2007ke does not have spectra at a sufficiently late epoch to test this.

On the other hand, if we require the absence of Hydrogen emission from this class, we would exclude
PTF\,09dav. It is the only member of this class that showed Hydrogen, albeit only at late times.

It is pertinent here to discuss the case of SN\,2008ha \citep{vpc+09,fcf+09,fbr+09}, a peculiar transient 
found in the disk of an irregular galaxy. This transient satisfies four out of the five properties of this class. 
The only difference is that it showed extremely low photospheric
velocities of $\approx$\,2000\,km\,s$^{-1}$. The photospheric spectra of SN\,2008ha have been
compared to those of the SN\,2002cx family of supernovae, and it has been suggested that SN\,2008ha is a lower velocity
and lower luminosity analog of this family. The SN\,2002cx family has starkly different late-time
spectra dominated by permitted iron lines (it has been argued that these spectra are not nebular even
at +400\,d, e.g., \citealt{jbc+06,sta+08}) and no evidence of Calcium-richness. Therefore, the 2002cx-family is likely unrelated to 
the class of transients discussed here. 

The membership of SN\,2008ha in either the SN\,2002cx family or in the ``Calcium-rich gap"  class
of transients is unclear. On grounds that the extremely low velocities may be diagnostic 
of a different explosion mechanism, we tentatively exclude SN\,2008ha as a member of the
class of ``Calcium-rich gap"  transients.

Originally, \citet{fcs+03} suggested that there was a class defined by only one property:
prominence of Calcium relative to supernovae of Type Ib. This property alone would suggest
that SN\,2000ds, SN\,2003dg, SN\,2003dr and SN\,2003H are also members
of this class. As discussed in \S\ref{sec:loss}, these events were not monitored photometrically.
Additionally, SN\,2005cz \citep{kmn+10,pgc+11}, which also has a similar nebular spectrum
but an extremely sparse light curve, would become a member of this class. 
Without a light curve, the ejecta mass in this classification
would remain ambiguous. As discussed below, low ejecta mass is a crucial clue to distinguish 
this class from variants of normal SN\,CC as well as SN\,Ia. We note here that even 
if light curves were available and consistent with the properties
of this class, it would not alleviate the challenge posed to many progenitor scenarios 
by the strong preference for remote locations. Specifically, SN\,2005cz is in an old environment
\citep{pgc+11}, SN\,2003H is amidst interacting galaxies, SN\,2000ds is in an elliptical galaxy,
and SN\,2003dr is offset from an edge-on galaxy.

%Finally, if we consider a class defined by location alone, say offset from host 
%greater than 30\,kpc, the class would include several normal SN\,Ia (discussed in \S2) 
%and two peculiar SN\,Ia (SN\,2006bt and PTF\,10ops; see \citealt{mst+11}). Clearly, the properties here
%are too diverse to represent a common explosion mechanism. Therefore, we do not a priori
%include location in the list of properties of this class.

\begin{deluxetable*}{llllll}%[!hbt]
  \tabletypesize{\footnotesize}
  \tablecaption{Summary of properties of ``Ca-rich gap'' transients}
  \tablecolumns{6}
  \tablewidth{0pc}
 \tablehead{\colhead{} & \colhead{SN\,2005E} & \colhead{SN\,2007ke} & \colhead{PTF\,09dav} & \colhead{PTF\,10iuv} & \colhead{PTF\,11bij}
}
 \startdata 
Peak absolute mag. ($M_{r}$)         & $-$15.5    &  $-$16.3   & $-$16.4    & $-$16      &  $-$15.9         \\
Rise-time (days)                      & $>$9   &  15     & 12      & 12      &  \nodata         \\
Decay time (days)                     & 14     &  15     & 12      & 12      &  10           \\ 
Photospheric velocity  (km\,s$^{-1}$)  & 11,000    & 11,000     & 6000      & 10,000    & \nodata         \\
Helium in photospheric spectra?       & Yes       &  Yes       & Maybe      & Yes        & \nodata         \\
Calcium in nebular spectra?           & Yes       & Yes        & Yes        & Yes        & Yes             \\
Hydrogen in nebular spectra?          & No        & No         & Yes        & No         & No              \\
Limiting magnitude of host galaxy & $-$7.5    & \nodata    & $-$9.8    & $-$12.1       & $-$12.4         \\
Offset from nearest host galaxy (kpc)       & 23   & 8.5   & 40    & 37     & 33         \\
Galaxy group environment?             & No        & Yes        & No         & Yes         & Yes                
 \enddata
\label{tab:summary}
\end{deluxetable*}

\begin{figure}[!htbp] 
   \centering
   \includegraphics[width=3.5in]{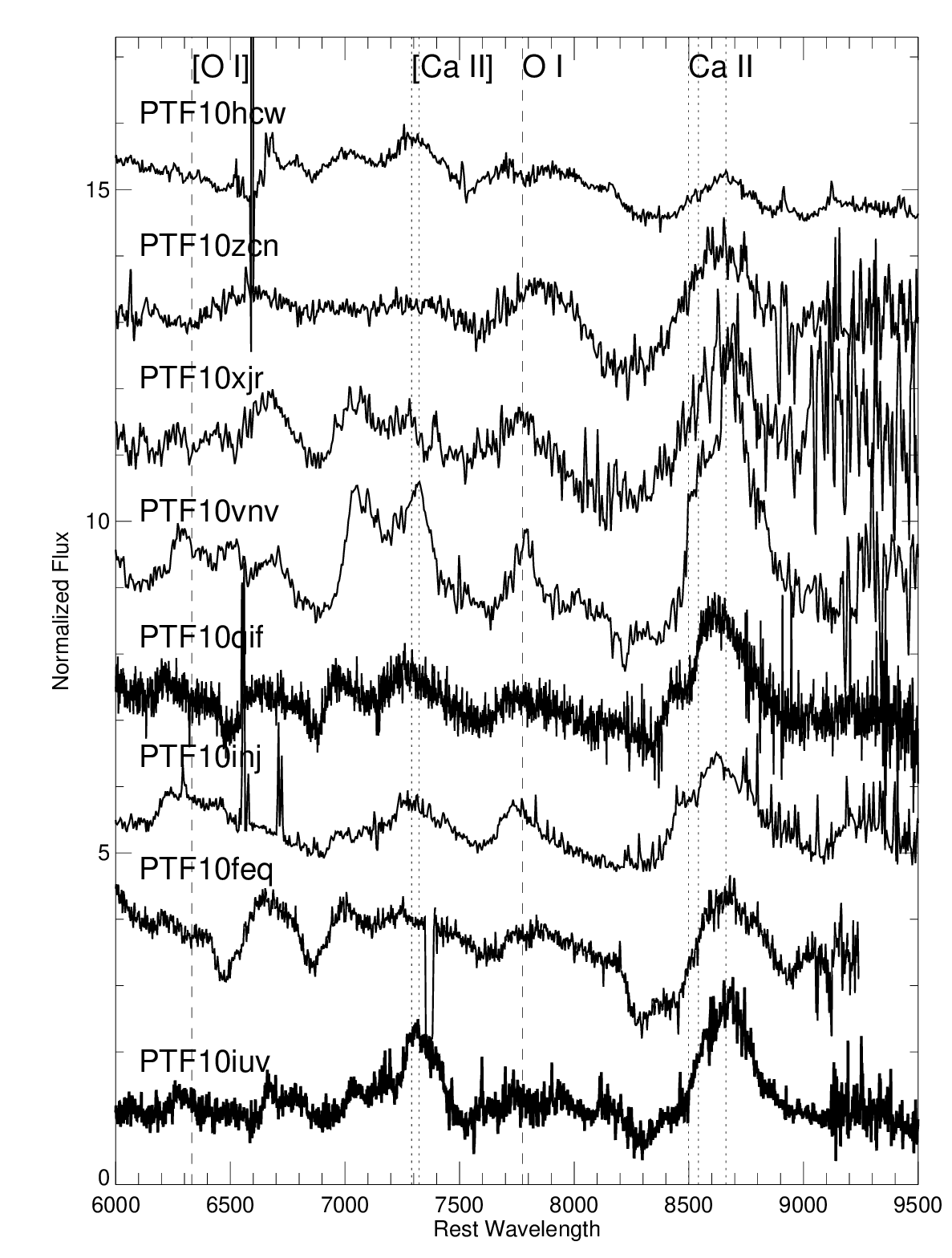}  
   \caption[Spectral Evolution of PTF10iuv]{\small Comparison of PTF\,10iuv
to other SN\,Ib about a month after maximum light. Note the strength of the [Ca~II] line 
relative to O~I. 
}
\label{fig:specIb}
\end{figure}

\subsection{Comparison to Type Ia and Type Ib Supernovae}
\label{sec:Iab}
In \citet{skn+11}, we discussed that although the photospheric spectra of PTF\,09dav
share some similarities to those of SN\,Ia, there are several differences as follows. 
(i) PTF\,09dav does not obey the Phillips relation \citep{p93}. There is a break in the slope of
the relation for subluminous SN\,Ia, but this is obeyed by all SN\,1991bg-like supernovae
including the faintest member, SN\,2007ax \citep{kog+08}. 
(ii) The photospheric spectra of PTF\,09dav show Scandium and Strontium 
(elements usually seen only in spectra of SN\,CC).
(iii) The nebular spectrum of PTF\,09dav does not show any Fe-peak elements,
which is inconsistent with nebular spectra of all SN\,Ia (Figure~\ref{fig:nebspec}).

The photospheric spectra of PTF\,10iuv around maximum light resemble those of SN\,Ib.
However, a distinguishing characteristic is that a month after maximum, Calcium emission 
starts to become prominent. We checked spectra of 26 SN\,Ib from PTF for 
Calcium-richness. In Figure~\ref{fig:spec10iuv}, we show seven spectra of SN\,Ib.
While the Calcium near-IR triplet becomes prominent in all of the SN\,Ib as they evolve,
the [Ca~II] doublet is especially prominent in PTF\,10iuv. PTF\,10vnv and PTF\,10inj also show
[Ca~II], but the relative flux ratio of [Ca~II] to O~I is much lower than that 
seen in PTF\,10iuv. The case of PTF\,10hcw is less clear, as [Ca~II] is more prominent
than O~I but the supernova set soon after discovery and the photometric coverage stops
before it reached maximum. In the nebular phase, the flux ratio of [Ca~II] to [O~I] is also 
much lower in SN\,Ib relative to PTF\,10iuv (Figure~\ref{fig:nebspec}). 
Furthermore, the light curves of SN\,Ib are typically more luminous by about 
2\,mag and evolve more slowly by at least a factor of 2.

\subsection{Modeling the Light Curve: Ejecta Mass and Radioactivity}
\label{sec:modellc}

\begin{figure}[htbp] 
   \centering
   \includegraphics[width=3.5in]{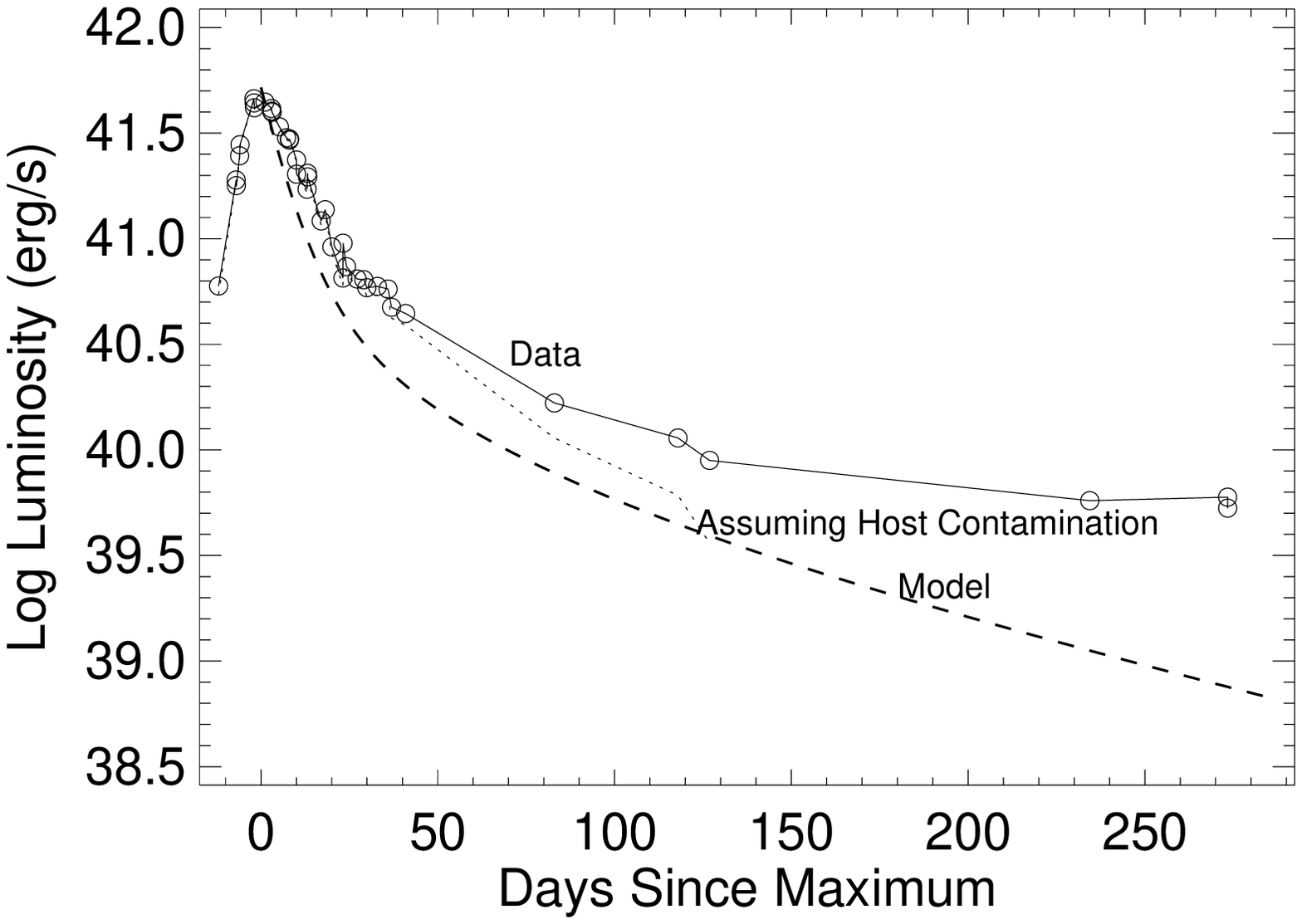} 
   \caption[Modeling the light curve of PTF10iuv]{\small Modeling the light curve 
(circles; $\nu\,F_{\nu}$ in the $r$ band) of PTF\,10iuv with radioactive $^{56}$Ni decay. The 
dotted line assumes that the last two data points are due to host-galaxy light contamination and 
subtracts $r=23.8$\,mag from the rest of the light curve. The model (dashed line) is based on an 
ejecta mass of 0.46\,M$_{\odot}$ and a $^{56}$Ni mass of 0.016\,M$_{\odot}$.}
\label{fig:bollc}
\end{figure}

The light curve of PTF\,10iuv is very well-sampled. 
The rising portion of the light curve can constrain 
the ejecta mass and the late-time decay can constrain the radioactive mass. 
Therefore, we can test the hypothesis of whether this
explosion is radioactively powered by $^{56}$Ni as in SN\,Ia.

For supernovae with photospheric velocity, $v$, and rise-time, $t_{\rm r}$,
the ejecta mass is $M_{\rm ej} \propto v\,t_{\rm r}^2$ \citep{a82}. 
Therefore, assuming the same opacity, we can derive an ejecta mass by scaling 
relative to a normal SN\,Ia with parameters 
1.4\,M$_{\odot}$, 11,000\,km\,s$^{-1}$, and 17.4\,days. 
PTF\,10iuv has a rise-time of 12\,days and an average 
photospheric velocity of 7600\,km\,s$^{-1}$; hence, it has 
an ejecta mass of 0.46\,M$_{\odot}$. PTF\,09dav has
the same rise-time but a lower velocity of 6000\,km\,s$^{-1}$,
giving an ejecta mass of 0.36\,M$_{\odot}$ \citep{skn+11}. There are no data to
constrain the rise-time of SN\,2005E. Assuming it was also 12\,days
and using its photospheric velocity of 11,000\,km\,s$^{-1}$, we 
get an ejecta mass of 0.67\,M$_{\odot}$. Note that this is a factor
of two higher than the total ejecta mass estimated by \citet{pgm+10} 
based on the nebular spectrum. This may not be surprising
given that the nebular spectrum analysis may not have accounted
for all of the Helium. Moreover, higher ejecta mass for SN\,2005E was
expected based on light-curve modeling as well \citep{wsl+10}.

For PTF\,10iuv, given the peak luminosity of $4.6\, \times \,10^{41}$\,erg\,s$^{-1}$ and a 12\,day rise, 
we can derive a $^{56}$Ni mass of 0.016\,M$_{\odot}$. However, we find that 
the late-time photometry is brighter than what is expected 
from radioactive decay of $^{56}$Ni (Figure~\ref{fig:bollc}).
A possible caveat here is that there is some light from an underlying dwarf
host galaxy. The last photometric point is fainter than the depth of our pre-explosion co-add. 
It is possible that there is some contamination in the light curve due to the
flux of such a host. Therefore, we plot the dotted line in Figure~\ref{fig:bollc} 
which subtracts the last two data points from the rest of the light curve. We are 
continuing to obtain even later time imaging to further test this caveat.

Should further observations rule out host contamination, 
this may suggest either a radioactive species other than $^{56}$Ni
or an additional source powering the light curve. 
For example, a larger contribution by $^{44}$Ti, which has a long half-life
of 60\,yr, may seem a promising candidate to explain the late-time slow evolution.
However, if all of the high luminosity at late times were attributed to this element, it 
would require too large a quantity of $^{44}$Ti ($\sim$\,2\,M$_{\odot}$). 
Modeling of different
combinations of other additional radioactive species ($^{44}$Cr, $^{44}$Ti,
and $^{52}$Fe) in the context of Helium-shell detonations
on small \citep{wsl+10} and large CO white dwarf cores \citep{skw+10} has been 
undertaken. These efforts were limited by the missing light-curve 
data in the rising and very late phase of SN\,2005E. The well-sampled
light curve of PTF\,10iuv should be able  to better constrain these models.

%There must atleast one other source powering the light curve.
%{\bf Photospheric velocities, the three Helium lines give 6524, 7649, 8727 km/s -- check}

%\subsubsection{Rise-time, Photospheric Velocity, Peak Luminosity}

%\begin{itemize}
%\item Any new elements based on SYNOW fits?
%\item Ratio of Calcium to Oxygen/Helium
%\item Equivalent Width of Calcium
%\item Velocity of Calcium
%\item Light Curve tau-mv plane
%\item Time to get nebular
%\item Time to see Calcium
%\item Offsets from Host
%\item Presence of [Ca II]?
%\end{itemize}

\subsection{Constraint on Electron Density}
Using the flux ratio between [Ca~II] and the Ca~II near-IR triplet,
we can constrain the density given a temperature. For PTF\,10iuv, 
the nebular spectrum at +87\,day gives a ratio of 0.86 and the 
spectrum at +115\,day gives a ratio of 1.8. For PTF\,09dav, 
the nebular spectrum at +94\,day gives a ratio of $>$2.2
(assumed width of 300\,\AA, 3$\sigma$ limit is 
$3 \times 10^{-18}$\,erg\,cm$^{-2}$\,s$^{-1}$\,\AA$^{-1}$).
For SN\,2005E, the ratio at +65\,d is 0.55 \citep{pgm+10}.
SYNOW fits to late-time spectra suggest a temperature
roughly in the range 4500--5000\,K. 
Following Figure~2 of \citet{fp89}, assuming a temperature
around 4500\,K, the electron density is on the order of $10^9$\,cm$^{-3}$
and decreases by a factor of a few in a couple of months.

\subsection{Modeling the Nebular Spectra}
Next, we estimate the mass of the dominant species in the 
ejecta using the nebular spectrum. 

We estimate the Oxygen mass based on the luminosity of the [O~I] line in 
the nebular phase. We can assume the high-density limit holds ($>10^{6}$\,cm$^{-3}$)
and estimate the oxygen mass as $$M_{O} = 10^{8}f_{\rm [O~I]} D_{\rm Mpc}^{2} e^{(2.28/T_4)} {\rm M}_{\odot},$$ following \citet{u86}.
Assuming a temperature of 4500\,K, we get 0.025\,M$_{\odot}$ of Oxygen
for PTF\,10iuv and 0.037\,M$_{\odot}$ for SN\,2005E. Note that the oxygen
mass is consistent with the mass derived by \citet{pgm+10}. A cautionary note
here is that this calculation is extremely sensitive to temperature; 
a difference of only 500\,K in temperature changes this estimate by a factor of 2.
%print, 97.^2*1.65e-16*1e8*exp(2.28/0.4)

Two months after maximum brightness, the luminosity in [Ca~II] nebular 
emission for SN\,2005E was $2 \times 10^{39}$\,erg\,s$^{-1}$ and the derived ejecta mass was 
0.135\,M$_{\odot}$ \citep{pgm+10}. The [Ca~II] nebular luminosity is a factor of 2.5 smaller for
PTF\,10iuv and factor of 2.6 larger for PTF\,09dav relative to SN\,2005E.
Under similar conditions, this may be representative of the range in Calcium mass
for these events.

\section{Discussion}
\label{sec:discussion}
%The answer to the fundamental question of whether the progenitor of this
%class of transients is a white dwarf or a massive star is not
%clear.

We are struck by the remote ``off-galaxy'' locations of all five confirmed
members of the class of ``Calcium-rich gap"  transients (Figure~\ref{fig:loc}) and
Figure~\ref{fig:lossfinders}). While five may still be considered small-number 
statistics, location can be a powerful diagnostic of the physical origin.
Below, we discuss the pros and cons of two broad classes of possible 
progenitor scenarios: a white dwarf and a massive star.

\subsection{A White Dwarf?}
%Pro: location, environment, low ejecta mass, similar LC
%Cons: Not a Ia, Not AIC, Not shell detonation
%
%All other lines of evidence suggest a white dwarf origin. The absence
%of a dwarf satellite galaxy host to deep limits for all three events, 
%combined with deep H$\alpha$ and ultra-violet limits for SN\,2005E, 
%suggests that a recent episode of in situ star formation is unlikely. 

The white dwarf scenario has tremendous appeal to explain ``Calcium-rich
gap'' transients for two reasons. 
First, the remote galactic locations, intracluster environments,
and constraints on star formation of the observed members of this family
point to an old population. Second, the low ejecta masses derived from
the light curves and nebular spectra are in the range of what is
derived from white dwarf explosions and not massive star explosions. 

The family of transients presented here is not the standard thermonuclear explosion
of a white dwarf as seen in SN\,Ia. There are two major
differences. First, none of these transients obey the \citet{p93} relation. The
inferred ejected masses are smaller than in SN\,Ia. 
Second, Fe-group elements seen in all SN\,Ia are absent in the nebular spectra of these explosions.
%Third, the presence of H$\alpha$ at late-time and Scandium and Strontium
%in early-time spectra of PTF\,09dav is also not seen in other SN\,Ia.
%with circumstellar medium. Second, the presence of Scandium and Strontium in
%the photospheric spectra of PTF\,09dav suggests conditions (e.g. lower temperature) 
%usually seen in core-collapse. Third, the absence of Fe-group elements seen in all 
%thermonuclear explosions and similarity to nebular spectra of core-collapse 
%(albeit with significantly enhanced Calcium) suggests
%this is a core-collapse. Fourth, the inconsistency of the light curve of PTF\,10iuv with 
%a radioactive $^{56}$Ni explosion and disobedience of the Phillips relation suggest
%this is not a regular Type Ia supernova.
%%similarity of photospheric spectra to Type Ib?

Next, we discuss several alternate explosions of white dwarfs that have been 
discussed in the literature to explain intermediate luminosity and fast
evolving transients. First, we consider a ``.Ia'' explosion following the final Helium flash in an ultracompact 
white-dwarf white-dwarf  binary \citep{bsw+07,skw+10}. The observed rise-time of ``Calcium-rich gap" 
transients is too slow for a ``.Ia'' explosion. The observed ejecta masses are too large for a 
shell detonation \citep{wsl+10} and suggest that the core also participated in the explosion.

\citet{wk11} derive several models for explosions of sub-Chandrashekhar white dwarfs.
However, the peak absolute magnitude and rise-time range observed for ``Calcium-rich gap''
transients is not predicted by any of these models.

Another proposed theoretical model is accretion induced collapse (AIC) of a rapidly rotating 
white dwarf into a neutron star \citep{mpq+09}. However, AIC predicts a spectrum dominated 
by intermediate-mass elements, much higher velocities (0.1\,$c$), and much more rapid rise 
(1\,day) and decline (4--5\,days) than what is observed for Ca-rich gap transients. 

Another possibility is deflagration of a sub-Chandrashekhar mass white dwarf \citep{ww94},
as it is also expected to have lower luminosity than a SN\,Ia. However, the late-time light curve 
and absence of Fe-peak elements in the nebular spectra are inconsistent with this model.

The presence of Helium in SN\,2005E and PTF\,10iuv can be explained for some 
white dwarf models. But, the presence of hydrogen in PTF\,09dav poses a major challenge to all white dwarf explosions.
One possible scenario to explain late-time hydrogen in a white-dwarf explosion is as follows:
Consider a binary where mass-transfer is from a hydrogen rich companion star onto a white dwarf. 
This accretion initially proceeded at a low rate, resulting in a series of nova eruptions prior to 
the sub-Chandrashekhar explosion or Helium-shell detonation. The photons from the final explosion would 
eventually reach one of the previously ejected nova shells and the interaction would give H$\alpha$ emission. 

%\subsection{Running into a Nova shell?}
%A possible scenario to explain the hydrogen from a white
%dwarf explosion is that this explosion was preceded by a nova-like event. 
%Only at late-time, the shock from the explosion reached the previously ejected 
%nova shell and interaction resulted in H$\alpha$ emission. 
Quantitatively, if the shell was photo-ionized, 
the distance to this nova shell would be the speed of light multiplied
by 95\,days, or $2.5 \times 10^{17}$\,cm. Given the velocity of 1000\,km\,s$^{-1}$
to traverse this distance, the nova eruption would then have occurred 78\,yr
prior to the supernova. Given the absence of Hydrogen in early-time spectra, we consider instead 
collisional ionization when the shock front itself reaches the shell. The 
distance to the shell would then be $8.2 \times 10^{15}$\,cm and the nova shell would
have to be ejected $\sim$\,950\,days before the supernova.  The mass of Hydrogen needed to sustain the observed 
luminosity for an hour is $\frac{4}{n}\,\times\,$10$^{-4}$\,M$_{\odot}$, where $n$ 
is the number of times the same Hydrogen atom gets excited in one hour. 

Yet another scenario was recently proposed by \citet{m11} that can  potentially
explain a trace amount of Hydrogen -- tidal disruption of a white dwarf by a neutron star 
or stellar mass black hole.  However, several other observables (e.g., velocities)
are inconsistent with this model.

%IGIMF
%the more massive a cluster, the higher the most massive star in it
%esp. in dwarf galaxies i.e. in the low H-alpha luminosity regime, the star formation 
%rate is much higher (as the fraction of massive stars is much lower) than that derived 
%by the Kennicutt relation

While the presence of five transients in remote locations is a strong clue in favor 
of the white dwarf scenario, the absence of any events in disks of galaxies is quite 
problematic. Studies of white dwarf binaries have shown that the number of white dwarfs
in halos relative to that in the disk is nearly a third, larger than expected from
the ratio of the overall mass or light \citep{rbb+07}. However, that is not sufficient
to explain why there is a strong preference for remote locations.

If the progenitor is indeed a white dwarf, then there must be a characteristic of white
dwarfs which can be attributed to location and affects the explosion signature.  
For example, the age of the white dwarf (e.g., if the time scale for evolution
in a multiple star system were longer), the composition of the white dwarf (e.g., if 
Helium white dwarfs were preferentially found in remote locations), environmental characteristics 
(e.g., if metallicity/density affected the explosion or progenitor evolution time scale).
One scenario that remains to be tested with detailed modeling is that
at lower metallicity, even relatively lower mass stars can evolve to become
white dwarfs in a Hubble time. Perhaps, explosions of these relatively 
lower mass white dwarfs are subluminous, fast evolving, and Calcium-rich.

\subsection{A Massive Star?}
Given that none of the above white dwarf models easily explain all of the observed
clues of ``Calcium-rich gap'' transients, we briefly consider the massive star option. There are 
some explosion properties which are reminiscent of core-collapse supernovae: similarity of nebular spectra
albeit with enhanced Calcium, similarity of photospheric
spectra to SN\,Ib in the case of PTF\,10iuv and SN\,2005E, 
and the presence of Hydrogen in the case of PTF\,09dav.

%Next, we begin with a discussion of whether massive stars can even be found in outskirts 
%and then discuss possible explosion mechanisms that may explain the low ejecta, low 
%luminosity and fast evolution.

%The biggest challenge to this hypothesis are the deep broad-band,
%H$\alpha$ and ultra-violet imaging limits against in situ star formation. 
%Next, we discuss the odds of finding a massive star in such remote locations.
%Assuming that this is possible, we discuss possible explosion channels.

\subsubsection{Massive Star in the Outskirts?}
\citet{pgm+10} convincingly argue against the progenitors being hypervelocity
massive stars formed in the disk. A more feasible option that allows massive stars
to traverse large distances is tidal stripping during galaxy interactions. 
This is especially intriguing given the galaxy group surrounding the locations 
of PTF\,10iuv, PTF\,11bij and SN\,2007ke.

Studies of intracluster environments show evidence for a dominant old component as well
as a young component, particularly in tidal tails of interacting gas-rich galaxies 
\citep{mkk+11,wcd+07}. It has been shown that roughly 15\% of the 
cluster's mass is in the intracluster medium.
%Williams et al. smallest age bin is 1 Gyr, so it is not a very good ref, lifetime goes as Mass^(-2.5)
Sivanandam et al. (2009) show that 50\% of metals come
from intracluster supernovae. \citet{mkk+11} present a study of tidal tails 
of interacting galaxies and find that young star clusters formed {\it in situ} in half the sample.  
Another possibility is that there is some low-level star formation in remote locations.
Recent systematic surveys  \citep{wpm+10,wpm+08} show star formation in the far outskirts 
of noninteracting galaxies. About 10\% of the galaxies
in this study had outlying H~II regions with offsets between 20\,kpc and 40\,kpc. 

Our deep ground-based limits show neither any evidence of tidal tails nor
H~II regions. However, given the distance to these supernovae, they are not
sensitive to a significant fraction of the luminosity function of H~II regions
and young star clusters. So, we cannot completely rule out the possibility of a massive 
star origin. Deep narrow-band imaging will provide better constraints. 
%<add limits including those for 05E and HII luminosity range>
%We note here that it has been argued that the maximum mass of a star formed in
%less massive clusters is lower. \citealt{kmn+10} discuss electron-capture
%induced collapse of an O-Ne-Mg core in extremely metal poor 6--10\,M$_{\odot}$
%stars. This may produce an explosion with intermediate luminosity 
%but not necessarily short duration. It is also quite challenging
%to have an IMF somehow truncated to 10\,M$_{\odot}$. 

\subsubsection{Fallback Supernova?}
Assuming there was a small amount of {\it in situ} star formation below our 
detection limits, one would expect regular core-collapse supernovae
in the outskirts as well. However, the absence of regular SN\,CC 
with offsets larger than 30\,kpc in the PTF sample appears to 
be at odds with the progenitors being massive stars (see Figure~\ref{fig:snoffsets}).

A possible partial resolution is if the fate of
massive stars in very low-metallicity environments is very different.
Specifically, the lower metallicity in the outskirts lowers the 
mass-loss rate, and it has been suggested that a larger fraction of massive stars 
collapse directly to a black hole (e.g., \citealt{hfw+03}, \citealt{oo11}). 
Such a collapse results in a subluminous explosion or no explosion at all. 
This could explain both the small numbers of SN\,CC
in the outskirts and the dearth of regular SN\,Ib/c
in low-metallicity environments. This is also consistent with studies that 
show that regular SN\,Ib/c are more centrally concentrated relative 
to SN\,II \citep{aj09}. Some of the missing Type Ib/c explosions in 
low-metallicity environments (outer parts of galaxies) could perhaps be
subluminous and short-lived fallback events. Fallback of some ejecta onto the proto-neutron star
to form the black hole could explain the low ejecta mass, fast evolution,
and absence of heavy elements observed in this class. It would not be surprising,
then, that the first fallback events observed are located in the outskirts of their hosts. 

There are three important caveats to this scenario. First, since the initial mass function is 
quite universal and varies little over a wide range of metallicity (e.g., \citealt{mkkm11}),
we expect to see a significant population of normal SN\,II-P from the lower 
mass range of SN\,CC unaffected by metallicity. But, current transient surveys 
have not found any SN\,II-P in such remote galactic outskirts. 
Second, PTF\,10iuv and SN\,2005E were Hydrogen-free. Therefore, mass loss
would have to be finely tuned to be high enough to expel the Hydrogen, low enough such that
there is fallback onto the core, yet not too low such that there is no explosion at all. 
\citet{kmn+10} proposed an alternate scenario where the Hydrogen was removed by close 
binary interaction. Third, some galaxies have shown 
that the metallicity gradient flattens beyond a certain radius and doesn't continue to decrease.
For example, metallicity measurements of the H~II regions in the outskirts 
show that they are not exceptionally low metallicity (roughly 0.4 solar; \citealt{wpm+11}). 
To the extent that metallicity is diagnostic of line-driven mass loss,
this presents another challenge to the massive star scenario.

%Looking into core-collapse sub-types, 
%recent studies \citep{aj09} have shown that Type Ib/c supernovae are 
%more centrally concentrated relative to Type II supernovae. \citealt{agk+10} also 
%show an underabundance of Type Ib/c relative to Type II in dwarf galaxies. Both these results show
%that Type Ib/c are only found in higher metallicity environments. In a recent review of progenitors
%of core-collapse, REF~S+09 show that Type II supernovae come from 
%progenitors in the mass range 10--25 M$_{\odot}$ and Type Ib/c from more massive
%progenitors. The absence of regular Ib/c in lower metallicity environments 
%raises the question of what is the stellar outcome of more massive stars in 
%the outer parts of the galaxy.
%
%{\bf This scenario has qualitative appeal. However, the catch is that the handful 
%of metallicity measurements of outlying HII regions doesn't suggest exceptionally low metallicity 
%(only 0.4 solar or so). Some galaxies show that the metallicity gradient flattens in the outskirts. 
%Considerable theoretical work would be needed to quantify how low a metallicity is needed to stop
%seeing regular Ib/c and start seeing fallback/direct collapse only. Alternately, if the nebular
%spectra or light curve can be modelled to derive some masses and the mass was greater than 1.4,
%that would settle the question.}
%
%{\it Note from Brad: Magnetar with 40 Msun progenitor in Westerlund I (Muno et al. 2006). 
%Check if this cluster especially offset or lower metallicity?}

\section{Conclusions}
\label{sec:conclusion}
%%Five with Five Properties. Caveats of larger/smaller sample.
%%Not standard white dwarf, Not standard core collapse
%%None of the proposed non-standard scenarios fit all the clues either
%%Considerable theoretical work is necessary
%%Observationally rates, stronger limits on star formation, systematic searches
%%and follow-up to determine whether or not location is a red herring

Five transients (SN\,2005E, SN\,2007ke, PTF\,09dav, PTF\,10iuv 
and PTF\,11bij) share the following common properties:
low peak luminosity, fast photometric evolution, large photospheric
velocities, early evolution to the nebular phase, and Calcium-dominated ejecta.
Furthermore, all five members of this class of ``Calcium-rich gap'' transients
are located in the outskirts of their 
putative host galaxies. This set of properties, in conjunction with peculiarities
specific to each of them (e.g., the presence of Hydrogen, Scandium,
and Strontium in PTF\,09dav, strong constraints against star 
formation in SN\,2005E, intracluster environment of PTF\,10iuv, PTF\,11bij, and SN\,2007ke),
warrants a creative modification of standard thermonuclear
or standard core-collapse scenarios. 

We can estimate a lower limit on the rate of this class of ``Calcium-rich gap'' 
events by comparing to the rate of SN\,Ia discovered by PTF in the same volume in the 
same time. Within 200\,Mpc, we found 128 SN\,Ia and 3 such events.
%%128 SN Ia and 138 SN II as of August 11, 2011 
Therefore, a lower limit on the rate is $>7 \times 10^{-7}$\,Mpc$^{-3}$\,yr$^{-1}$
(we used the SN\,Ia rate from \citealt{lcl+11}). We emphasize that this 
is a lower limit, as a few days of bad weather would be much more detrimental for
finding these short-lived and lower luminosity transients compared to SN\,Ia. 
This rate of $>$2.3\% relative to SN\,Ia is consistent with the 
relative rate of 7$\pm$5\% estimated by \citet{pgm+10}. 

%LAST PARA?
The key to solving the mystery of the origin of ``Calcium-rich gap''
transients is their location. The remote locations can either be
a red herring (due to small numbers) or the most important clue. Parenthetically, we 
remark that the two Type Ia supernovae (SN\,2006bt -- \citealt{fnc+10}; 
PTF\,10ops -- \citealt{mst+11}) found farthest outside the main body 
of their host galaxies are also peculiar.
Extremely deep late-time observations (e.g., with the {\it Hubble Space Telescope}) 
are needed to serve as a litmus test: detection of a dwarf satellite,
super star cluster, or globular cluster would provide a direct clue
about the environment, and nondetection would require a phoenix 
transient that rose from the ashes.

%OLD LAST PARA:
%The way forward to ascertain the nature of the explosion of
%Calcium-rich gap transients requires both theoretical and
%observational work. Progress theoretically requires studying 
%alternative channels of white dwarf explosions that satisfy 
%all the observables and have a preference for remote locations.
%Specifically, further exploring the scenario of longer evolution 
%time scale of lower mass white dwarfs and their subsequent explosion
%properties appears promising. Alternately, quantitative modelling of the 
%metallicity dependence in core-collapse supernovae would help better 
%understand whether the lower end of the mass range may also be affected. 
%Progress observationally requires a larger sample of discoveries that 
%are intensely followed up, with special attention to obtaining late-time 
%nebular spectra, deeper constraints on in situ star formation and extensive
%environmental studies. This would answer the question of whether the remote 
%location is a red herring or the most important clue. Fortunately, ongoing 
%synoptic surveys such as the Palomar Transient Factory are well-poised to 
%uncover and follow-up new members of this class. \\

\bigskip
\smallskip
\noindent{\bf Acknowledgments} \\
We thank Chuck Steidel and Ryan Trainor for Target of Opportunity 
observations of PTF\,11bij with Keck I. We are grateful to the staff of 
the Expanded Very Large Array for efficiently executing Target Of 
Opportunity triggers. Excellent assistance was provided by the staffs 
of the various observatories where we obtained data (Lick, Keck, Palomar,
WHT, EVLA).

MMK acknowledges support from the Hubble Fellowship and Carnegie-Princeton
Fellowship. EOO is supported by an Einstein Fellowship. HBP is a 
CfA and Bikura prize fellow.
The Weizmann Institute's participation in PTF is supported by grants
to AGY from the Israel Science Foundation and the US-Israel Binational
Science Foundation. AVF and his group at UC Berkeley acknowledge generous financial
assistance from Gary \& Cynthia Bengier, the Richard \& Rhoda
Goldman Fund,
the TABASGO Foundation, and National Science Foundation (NSF) grant AST-0908886. 
KAIT was constructed and supported by donations from Sun
Microsystems, Inc., the Hewlett-Packard Company, AutoScope
Corporation, Lick Observatory, the NSF,
the University of California, the Sylvia \& Jim Katzman
Foundation, and the TABASGO Foundation.
Computational resources and data storage were contributed by NERSC, supported by 
U.S. DoE contract DE-AC02-05CH11231. PEN acknowledges support from the US DoE 
contract DE-FG02-06ER06-04. JSB acknowledges support of an NSF-CDI grant 0941742, 
``Real-time Classification of Massive Time-series Data Streams.''
MS acknowledges support from the Royal Society. LB acknowledges NSF grants
PHY-0551164 and AST-0707633. DCL acknowledges NSF grant AST-1009571.

Some of the data presented herein were obtained at the W. M. Keck Observatory, which is operated 
as a scientific partnership among the California Institute of Technology, the University 
of California and the National Aeronautics and Space Administration; 
it was made possible by the generous financial support of the W. M. Keck Foundation.
The Expanded Very Large Array is operated by the National Radio Astronomy
Observatory, a facility of the NSF operated
under cooperative agreement by Associated Universities, Inc.
The WHT is operated on the island of La Palma by the Isaac Newton Group in the Spanish
Observatorio del Roque de los Muchachos of the Instituto de Astrofisica de Canarias.

\bibliographystyle{apj}
\bibliography{ms}

\end{document}